\documentclass[journal]{IEEEtran}
\usepackage{times,amsmath,amssymb,amsfonts}
\usepackage{amsthm}
\usepackage{graphicx,epsfig,psfrag,color,cite}
\usepackage{algorithm}
\usepackage{algorithmic}

\def\HOME{D:/Dropbox/}
\def\bibhome{\HOME/Research/Publications/Bibs}

\IEEEoverridecommandlockouts

\def\reals{ { {\rm  I \kern-0.15em R }  } }
\def\complex{ {\,{{\rm C} \kern-0.50em \raise0.20ex {  |}}\, }}
\def\alphabf{\boldsymbol \alpha}

\def\gammabf{\boldsymbol \gamma }
\def\deltabf{\boldsymbol \delta }

\def\lambdabf{\boldsymbol \lambda }

\def\nubf{\boldsymbol \nu }

\def\Deltabf{\boldsymbol \Delta}

\def\Lambdabf{\boldsymbol \Lambda}

\def\abf{{\bf a}}
\def\bbf{{\bf b}}
\def\cbf{{\bf c}}

\def\fbf{{\bf f}}
\def\gbf{{\bf g}}
\def\hbf{{\bf h}}

\def\qbf{{\bf q}}

\def\ubf{{\bf u}}
\def\vbf{{\bf v}}
\def\wbf{{\bf w}}
\def\xbf{{\bf x}}

\def\zbf{{\bf z}}

\def\xbf{{\bf x}}

\def\Abf{{\bf A}}

\def\Cbf{{\bf C}}
\def\Dbf{{\bf D}}

\def\Gbf{{\bf G}}
\def\Hbf{{\bf H}}
\def\Ibf{{\bf I}}

\def\Rbf{{\bf R}}

\def\Ubf{{\bf U}}
\def\Vbf{{\bf V}}

\def\Xbf{{\bf X}}

\def\Cc{{\cal C}}
\def\Dc{{\cal D}}

\def\Gc{{\cal G}}

\def\Kc{{\cal K}}
\def\Lc{{\cal L}}

\def\Nc{{\cal N}}
\def\Oc{{\cal O}}
\def\Pc{{\cal P}}

\def\Sc{{\cal S}}

\theoremstyle{definition}

\newtheorem{theorem}{Theorem}
\newtheorem{corollary}{Corollary}
\newtheorem{lemma}{Lemma}

\newtheorem{proposition}{Proposition}

\IEEEoverridecommandlockouts
\DeclareMathOperator{\diag}{diag}
\renewcommand\Re{\operatorname{\mathfrak{Re}}}

\def\nn{\nonumber}
\def\defeq{\triangleq}
\def\eg{{\it e.g.,} }

\def\ie{{\it i.e.,\ \/}}
\def\tr{{\rm tr}}

\def\st {\text{subject~to~}}

\def\tot{\text{tot}}

\def\SCA{\text{\tiny SCA}}
\def\SINR{\textrm{SINR}}
\def\MMF{\text{\tiny MMF}}
\def\QoS{\text{\tiny QoS}}

\begin{document}

\title{Multi-group Multicast Beamforming: Optimal Structure and Efficient Algorithms}

\author{Min Dong\thanks{The authors are with
the Department of Electrical, Computer and Software Engineering, Ontario Tech University, Ontario L1G 0C5, Canada. Email: \{min.dong, qiqi.wang\}@ontariotechu.ca. Part of  this work was presented in \cite{Dong&Wang:SPAWC19}. }, \IEEEmembership{Senior Member,~IEEE}, and Qiqi Wang}
\markboth{}{}
\maketitle

\begin{abstract}
This paper considers the multi-group multicast beamforming optimization problem, for which the optimal solution has been unknown due to the non-convex and NP-hard nature of the problem. By utilizing the successive convex approximation numerical method and Lagrangian duality, we obtain the optimal multicast beamforming solution structure for both the quality-of-service (QoS) problem and the max-min fair (MMF) problem. The optimal structure brings valuable insights into multicast beamforming: We show that the notion of uplink-downlink duality can be generalized to the multicast beamforming problem. The optimal multicast beamformer is a weighted MMSE filter based on a group-channel direction: a generalized version of the optimal downlink multi-user unicast beamformer. We also show that there is an inherent low-dimensional structure in the optimal multicast beamforming solution independent of the number of transmit antennas, leading to efficient numerical algorithm design, especially for  systems with large antenna arrays. We propose efficient algorithms to compute the multicast beamformer based on the optimal beamforming structure.  Through asymptotic analysis, we characterize the asymptotic behavior of the multicast beamformers as the number of antennas grows, and in turn, provide simple closed-form approximate multicast beamformers for both the QoS and MMF problems. This approximation offers practical multicast beamforming solutions with a near-optimal performance at very low  computational complexity for large-scale antenna systems.
\end{abstract}

{\IEEEkeywords Multicast beamforming, optimal solution structure, duality, large-scale antenna systems, computational complexity, asymptotic beamforming
}

\section{Introduction}
We consider the  downlink multi-group multicast beamforming problem. In wireless communications, common data may be intended for a group of users.   Multi-antenna multicast beamforming is an efficient physical-layer transmission technique to deliver  common data to multiple users simultaneously, improving both spectrum and power efficiency. Multicast transmit beamforming has been first considered more than a decade ago  \cite{Sidiropoulosetal:TSP06}. The attention to this technique is fast rising  in recent years for its potential to support  wireless multicasting and content distribution in the growing number of    wireless services and applications  (\eg video conference, mobile commerce, intelligent transportation systems).
Besides these, in the emerging    cache-aided wireless networking   technologies,  (coded) multicasting is utilized in coded caching techniques for   content delivery of individual data requests to reduce wireless traffic
 \cite{Maddah-Ali&Niesen:TIT2014}.
 This new area of application further expands the potential of  multicast beamforming techniques in improving content distribution and delivery  in the rising trend of content-centric wireless networks.

The problem of multicast beamforming  optimization  has initially been considered for a single user group \cite{Sidiropoulosetal:TSP06,WuMaSo:TSP13,Abdelkaderetal:TSP10}. It has  later been extended to multiple user groups \cite{Karipidisetal:TSP08,ChangLuoChi:TSP08,Ottersten&etal:TSP14} 
  and multi-cell networks \cite{Jordan&Gong&Ascheid:Globecom09,Xiang&Tao&Wang:IJWC:13},  where inter-group or inter-cell interference  further complicates the problem.
Two types of problem formulation are typically considered for multicast beamforming: the transmit power minimization  subject to a minimum signal-to-interference-and-noise (SINR)  target for each user -- the quality of service (QoS)  problem, and the maximization of (weighted) minimum SINR of all users subject to a total transmit power budget -- the max-min fair (MMF) problem.
The family of these  multicast beamforming problems are non-convex and are shown to be NP-hard in general \cite{Sidiropoulosetal:TSP06}. Existing literature works have focused on developing numerical algorithms or  signal processing methods to obtain feasible solutions with good performance.
It is more direct to solve the QoS problem than the MMF problem, although the  feasibility  of the QoS problem imposes challenges in designing numerical methods. The MMF problem is typically handled  in the literature works by iteratively solving the QoS problem. 

Among existing methods for tackling the multicast beamforming problems, semi-definite relaxation (SDR) is a popular numerical approach to obtain an approximate (sometimes global optimal) solution  by relaxing the problem into a semi-definite problem (SDP) to solve \cite{Sidiropoulosetal:TSP06,WuMaSo:TSP13,Karipidisetal:TSP08,Luo&etal:SPMag10,Xiang&Tao&Wang:IJWC:13}.  Its provable approximation accuracies  are shown via theoretical analysis \cite{Luoetal:SIAM07}. However, as the problem size increases, the computational complexity of SDR-based methods grows quickly,  and the  performance deteriorates noticeably \cite{Karipidisetal:TSP08}. These drawbacks make the direct use of this approach unsuitable for     future large-scale wireless systems, in particular for massive multiple-input and multiple-output (MIMO) systems with large-scale antenna arrays   \cite{Rusek&etal:SPM13}\footnote{The computational complexity to solve the QoS problem directly via SDR is $\Oc(N^6)$ where $N$ is the number of antennas at the base station.}. For addressing these issues, the successive convex approximation (SCA) \cite{Marks&Wright:OR78}  has been proposed for multicast beamforming   in large-scale systems  \cite{LeNam&etal:SPL14,Christopoulos&etal:SPAWC15,Scutari&etal:SP17}. The SCA is an iterative numerical  method to solve the  original problem through a sequence of convex approximations. Although it shows good performance with  reduced complexity, the SCA requires an initial feasible  solution  for the problem that is generally difficult to obtain. Besides, its computational complexity is still high for a large number of antennas. There is an increasing need for effective and efficient multicast beamforming design. To further address the computational complexity,  low-complexity multicast
beamforming schemes  have  recently been proposed   for massive MIMO systems for  multi-group \cite{Sadeghi&etal:TWC17,Chen&Tao:TCOM17} and multi-cell \cite{Yu&Dong:ICASSP18, Yu&Dong:SPAWC18} scenarios. These schemes use specific beamforming strategies (\eg maximum ratio transmission (MRT) or zero-forcing (ZF)) in combination with SCA or distributed optimization techniques to reduce  complexity.

A primary challenge for the multicast beamforming problems is the elusive optimal solution structure. Prevailing numerical optimization methods   target at finding  good feasible solutions to the non-convex problems. However, theoretically, they are unable to characterize or  offer a fundamental understanding of the beamforming structure for multicasting, and practically, they face challenges in both computational complexity and performance  for large-scale systems. In this paper, in the general multi-group setting, we characterize the optimal multicast beamforming  structure for  both QoS and MMF problems. We take a  different approach from existing works by exploring both the numerical method of iterative approximation via SCA and Lagrangian duality  and combining the two techniques to obtain  the optimal multicast beamforming solution structure for the QoS problem. This solution structure   provides valuable  insights into the optimal multicast beamforming: We establish  an uplink-downlink duality interpretation for downlink multicast beamforming, as a generalization of the uplink-downlink duality for downlink multi-user unicast beamforming. We show that the optimal  beamforming solution for a multicast group has an intuitive structure: a weighted minimum mean square error (MMSE) filter, formed by the group-channel  direction  and the noise plus weighted channel covariance matrix. This optimal multicast beamforming is  a generalized version of the optimal downlink multi-user unicast beamforming. We draw connections and explain differences between the multicast and unicast beamforming.
An important finding from the optimal structure  is that the optimal multicast beamforming has an inherent low-dimensional structure, where  only weights of  user channels in the group need to be computed. This structure changes the  dimension of the multicast beamforming problem from the number of antennas to the number of users per group, and the problem size   may be further reduced depending on the dimension of the subspace spanned by the  user   channels in each group. This  optimal multicast beamforming structure gives rise to  efficient  numerical methods to compute the beamforming solution, especially for  massive MIMO systems with large-scale antenna arrays.

With the optimal  multicast beamforming structure, we only need to compute the parameters in the optimal   solution. Due to the NP-hard nature of the original problem, obtaining their optimal values are difficult. We propose efficient numerical algorithms to compute these parameters, including  Lagrange multipliers (associated with the SINR constraints) and user weights (in the group-channel direction). Our algorithm for computing the Lagrange multipliers is asymptotically optimal. We derive   the  asymptotic expression of the multipliers. It can be used directly for systems with a large number of antennas, further eliminating the computational need. To compute the weights, we transform the original problem into a weight optimization problem of a much smaller size, independent of the number of antennas. Taking advantage of this small  problem size, we apply the SDR or SCA method for good approximate or locally optimal solutions with very low computational complexity.

We extend our results to the MMF problem. Exploring the inverse relation of the QoS and MMF problems, we directly obtain the optimal MMF multicast beamforming structure. Computing the MMF beamformers is more involved, which requires iteratively computing the QoS beamformers. However,  we show that the asymptotic results obtained from the QoS problem lead to  simple   asymptotic  MMF beamformers, including a closed-form asymptotic beamformer. They provide simple   approximate multicast beamforming solutions for  a large number of antennas. Simulation demonstrates both the computational efficiency  and the near-optimal performance by our proposed algorithms using the optimal multicast beamforming structure.

\subsection{Related Work}
Downlink multicast transmit beamforming  has been studied for both QoS and MMF problems in  single-group \cite{Sidiropoulosetal:TSP06,Abdelkaderetal:TSP10,Wenetal:SAM12,WuMaSo:TSP13,KimLovePark:TCOM11} and multi-group \cite{Karipidisetal:TSP08,ChangLuoChi:TSP08,Ottersten&etal:TSP14} settings, as well as in multi-cell scenarios \cite{Jordan&Gong&Ascheid:Globecom09,Xiang&Tao&Wang:IJWC:13}. It has also been considered in other network scenarios, such as relay networks \cite{Bornhorstetal:TSP11,DongLiang:CAMSAP13}, cognitive spectrum access \cite{phan&etal:TSP09}, and cache-aided cloud radio access networks \cite{Tao&Chen&Zhou&Yu:ITWC16}.   
The family of problems are non-convex quadratically
constrained quadratic programming (QCQP) problems and are shown to be NP-hard in general \cite{Sidiropoulosetal:TSP06}. The SDR approach was proposed \cite{Sidiropoulosetal:TSP06} and has been widely used \cite{WuMaSo:TSP13,Karipidisetal:TSP08,Xiang&Tao&Wang:IJWC:13,Ottersten&etal:TSP14}, due to its bounded approximation performance \cite{Luoetal:SIAM07} and can be efficiently solved by interior-point methods  with polynomial time complexity \cite{Boyd:book} for  problems with a moderately small size.
Different techniques have been proposed to extract a rank-one approximate solution to the original problem from the relaxed problem, including randomization methods \cite{Sidiropoulosetal:TSP06,WuMaSo:TSP13} and rank-reduction methods \cite{HuangPalomar:TSP10}. The conditions for the existence of an optimal rank-one solution for the SDR problem were also investigated \cite{HuangPalomar:TSP10}.
Rank-two multicasting beamforming techniques were also proposed as a generalization  of the rank-one SDR-based approach by combining beamforming and the Alamouti space-time code \cite{WuMaSo:TSP13,Wenetal:SAM12}. Alternative signaling processing   approaches, such as channel orthogonalization, were also proposed \cite{Abdelkaderetal:TSP10,KimLovePark:TCOM11}.

Recently, a great deal of effort has been made in developing  computationally efficient numerical methods for massive MIMO systems with large-scale antenna arrays \cite{LeNam&etal:SPL14,Xiang&Tao&Wang:IJSAC14,Christopoulos&etal:SPAWC15,Sadeghi&etal:TWC17,Chen&Tao:TCOM17,Scutari&etal:SP17,Yu&Dong:ICASSP18,Yu&Dong:SPAWC18}. The SCA method is applied to  find a stationary solution for single-group \cite{LeNam&etal:SPL14}, multi-group \cite{Christopoulos&etal:SPAWC15}, and multi-cell \cite{Scutari&etal:SP17} scenarios. It  is shown to perform better than SDR-based methods in large-scale systems with reduced computational complexities. However, the SCA method requires a feasible initial point  that is difficult to obtain  in general. Several optimization techniques have been developed to improve SCA-type methods \cite{Mehanna&etal:SPL15,Scutari&etal:SP17}. For massive MIMO systems,  the existing SCA-based methods are still computationally intensive. Asymptotic multicast beamformers were derived by invoking channel orthogonality   at the asymptotic regime to eliminate interference \cite{Xiang&Tao&Wang:IJSAC14,Sadeghi&Yuen:Globecom15}. While they have simple analytical expressions, it is observed that these beamformers   perform poorly in most practical systems \cite{Sadeghi&Yuen:Globecom15,Yu&Dong:ICASSP18}. We will explain this phenomenon  of slow convergence  to asymptotic orthogonality in Section~\ref{subsubsec:AS} through our asymptotic analysis.  Several low-complexity methods have been proposed for massive MIMO systems. These include   a two-layer method combining ZF and SCA \cite{Sadeghi&etal:TWC17} and an alternating direction method of multipliers (ADMM)  algorithm \cite{Chen&Tao:TCOM17} for multi-group multicasting, and  a weighted MRT beamforming method  for both centralized and distributed coordinated multicast beamforming in multi-cell scenarios \cite{Yu&Dong:ICASSP18, Yu&Dong:SPAWC18}.

Besides the above, multicast beamforming in overloaded systems with fewer antennas than the users  has also been investigated recently. With insufficient degrees of freedom,  the rate-splitting  based   MMF beamforming strategies have been proposed for single-carrier or multi-carrier systems \cite{Joudeh&Clerckx:WCOM17,Chen&etal:TVT20,Tervo&etal:SPAWC18}.
Multicast beamforming using other design objectives has also
been considered, such as energy efficiency maximization  \cite{Tervo&etal:SPAWC18,Tervo&etal:SP18} and  the sum-rate maximization in a mixed
multicast and broadcast scenario  \cite{Yalcin&Yuksel:COM19}.
  
\subsection{Organization and Notations}
The rest of this paper is organized as follows. In Section~\ref{sec:system}, we present the system model and problem formulation for multi-group multicast beamforming. In Section~\ref{sec:opt_solution}, we  derive our main result of the optimal multicast beamforming in semi-closed-form for the QoS  problem and characterize the solution structure. In Section~\ref{sec:num}, numerical algorithms are proposed for computing the parameters in the optimal solution, and asymptotic analysis is provided at the large-scale antenna array regime.  In Section~\ref{sec:MMF}, we describe the optimal solution structure for the MMF problem, its relation to the solution structure for the QoS problem, and the asymptotic MMF beamforming solution. Simulation results are presented in Section~\ref{sec:sim}, and the conclusion and possible extension are provided in Section~\ref{sec:con}.

\emph{Notations}: Hermitian, transpose, and conjugate are denoted as $(\cdot )^H$, $(\cdot)^T$, and $(\cdot)^*$, respectively.  The Euclidean
norm of a vector is denoted by $\left\|\cdot\right\|$.
The notation  $\abf \succcurlyeq {\bf 0}$ means element-wise  non-negative, and $\Abf \succcurlyeq  0$ indicates  matrix $\Abf$ being positive  semi-definite.
The trace of matrix $\Abf$ is denoted as $\tr(\Abf)$.  The real part of $x$ is denoted by $\Re\{x\}$, and $E(x)$ denotes the expectation of $x$. The abbreviation i.i.d. stands for independent and identically distributed,
and $\xbf \sim \Cc\Nc({\bf 0}, \Ibf)$ means $\xbf$ is a complex Gaussian random vector with zero mean and covariance $\Ibf$.

\allowdisplaybreaks

\section{System Model and Problem Formulation} \label{sec:system}
Consider a downlink multi-group multicasting scenario, where a BS equipped with $N$ antennas serves $G$ multicast groups, sending  each group a common message that is independent of other groups. Let $\Gc=\{1,\ldots,G\}$ denote the set of  group indices. Each group $i$ consists of $K_i$ single-antenna users, and the set of user indices in the group is denoted by $\Kc_i=\{1,\ldots,K_i\}$,   $i\in \Gc$.  User groups are disjoint, \ie each user  only belongs to one multicast group and only receives the multicast message intended to this group. The total number of users in all groups is denoted by $K_\tot\triangleq\sum_{i=1}^G K_i$.

Let $\hbf_{ik}$ denote the $N\times 1$ channel vector between the BS and user $k$ in group $i$,  and let $\wbf_i$ denote the $N\times 1$ multicast beamforming vector for  group   $i \in \Gc$. The received signal at user $k$ in group $i$ is given by
\begin{align} \label{y_mk}
y_{ik} = \wbf_i^H\hbf_{ik}s_i +  \sum_{j\neq i} \wbf_{j}^H\hbf_{ik}s_j + n_{ik}
\end{align}
where $s_i$ is the data symbol intended for group $i$ with unit power $E(|s_i|^2)=1$, and $n_{ik}$ is the receiver additive white Gaussian noise  with zero mean and variance $\sigma^2$.  The transmit power at the BS is given by $\sum_{i=1}^G \|\wbf_i\|^2$. The received SINR at user $k$ in group $i$ is given by
\begin{align} \label{multi:SNR}
\SINR_{ik} &= \displaystyle  \frac{|\wbf_i^H\hbf_{ik}|^2}{\displaystyle \sum_{j\neq i}  |\wbf_j^H\hbf_{ik}|^2+\sigma^2}.
\end{align}

Depending on the design focus, two problem formulations are typically considered for the multicast beamforming: 1) the QoS problem for transmit power minimization while meeting the received SINR target at each user,   formulated as \begin{align}
\Pc_o: \  \min_{\wbf}  & \ \ \sum_{i=1}^G \|\wbf_i\|^2 \nn \\
\st & \ \ \SINR_{ik} \ge \gamma_{ik}, \ k\in \Kc_i, ~i \in \Gc \label{SINR_constr}
\end{align}
where   $\wbf\triangleq[\wbf_1^H,\ldots,\wbf_G^H]^H$, and  $\gamma_{ik}$ is the SINR target at user $k$ in group $i$.  2) The (weighted) MMF problem for maximizing the minimum (weighted) SINR, subject to the transmit power constraint, formulated as
\begin{align}
\Sc_o: \  \max_{\wbf}\min_{i,k}  & \ \ \frac{\SINR_{ik}}{\gamma_{ik}}
\nn \\
\st  &\ \   \sum_{i=1}^G \|\wbf_i\|^2 \le P \nn
\end{align}
where $P$ is the transmit power budget, and $\{\gamma_{ik}\}$ here serve as the  weights to control the fairness or service grades among users.

\vspace*{0.5em}
\noindent{\bf Remark} (\emph{Feasibility}):
 The QoS problem $\Pc_o$ for multi-group multicast beamforming may not  always be feasible, depending on the channels $\{\hbf_{ik}\}$ and the SINR targets $\{\gamma_{ik}\}$. 
On the other hand, the MMF problem $\Sc_o$ is always feasible, but more involved than the QoS problem to solve. In the following sections, we assume the QoS problem $\Pc_o$ being feasible to derive the optimal multicast beamformer structure.

\section{Optimal Multicast Beamforming Structure }\label{sec:opt_solution}

We now focus on the multicast beamforming QoS problem $\Pc_o$, which is known to be a non-convex QCQP problem and  NP-hard. The optimal solution is difficult to obtain either in the primal  domain, or in the dual domain due to the unknown    duality gap. In the following, we take a different approach by exploring the problem via the SCA method  and derive the structure of the optimal  solution.

\subsection{The SCA Method}\label{subsec:SCA}
The SCA method is a numerical approximation method  that iteratively solves a non-convex optimization problem via a sequence of convex approximations of the original problem, provided that an initial feasible point is given. For non-convex problems with a convex objective function, the SCA method is proven to converge to a stationary solution \cite{Marks&Wright:OR78}. The SCA method, and in particular, the convex-concave procedure (CCP) as a special case,  has been applied to find a feasible multicast beamforming solution in several existing works \cite{LeNam&etal:SPL14,Christopoulos&etal:SPAWC15,Scutari&etal:SP17,Sadeghi&etal:TWC17}. The SCA method is briefly described below.

Consider $N\times 1$ auxiliary vector $\zbf_i$, $i \in \Gc$. For matrix  $\Abf\succcurlyeq {\bf 0}$, we have $(\wbf_i-\zbf_i)^H\Abf(\wbf_i-\zbf_i)\ge 0$, for any $\zbf_i$. It follows that
$\wbf_i^H\Abf\wbf_i\ge
2\Re\{\wbf_i^H\Abf\zbf_i\}-\zbf_i^H\Abf\zbf_i$. Denote $\zbf \triangleq [\zbf_1^H,\ldots,\zbf_G^H]^H$.
Given $\zbf$, applying the above inequality to SINR\ constraint \eqref{SINR_constr}, we obtain the following optimization problem which is a convex approximation  of $\Pc_o$
\begin{align}
\hspace*{-.2em}\Pc_\SCA(\zbf):  \min_{\wbf} & \  \sum_{i=1}^G \|\wbf_i\|^2 \nn \\
\st & \ \gamma_{ik}   \sum_{j\neq i}  |\wbf_j^H \hbf_{ik}|^2 -2\Re\{\wbf_i^H\hbf_{ik}\hbf_{ik}^H\zbf_i\} \nn \\
& \ \ +|\zbf_i^H\hbf_{ik}|^2 \le-\gamma_{ik}\sigma^2, \ k\in \Kc_i, i \in \Gc. \label{SCA_constr}
\end{align}
With  non-convex SINR constraint \eqref{SINR_constr} being replaced by  convex constraint  \eqref{SCA_constr}, problem $\Pc_\SCA(\zbf)$ is now convex. The main steps in the SCA method are summarized below:
\begin{enumerate}
\item Set initial feasible point $\zbf^{(0)}$; Set $l=0$.
\item  Solve  $\Pc_\SCA(\zbf^{(l)})$ and obtain the optimal solution $\wbf^\star(\zbf^{(l)})$.
\item Set $\zbf^{(l+1)}=\wbf^\star(\zbf^{(l)})$.
\item Set $l \leftarrow l+1$. Repeat Steps 2-4 until convergence.
\end{enumerate}

The above  SCA method is guaranteed to converge to a stationary point $\wbf^\star$ of $\Pc_o$ \cite{Marks&Wright:OR78}. Since the global optimal solution is  a stationary point, the above procedure may converge to the global optimal solution $\wbf^o$ of $\Pc_o$, provided that the initial point $\zbf^{(0)}$ is appropriately chosen, \eg $\zbf^{(0)}$ is at the vicinity of $\wbf^o$. When this is the case, we have $\zbf^{(l)} \to \wbf^\star =\wbf^o$.

\vspace{0.5em}
\noindent{\bf Remark}: A challenge to use the SCA method for $\Pc_o$ is  finding an initial feasible point $\zbf^{(0)}$ that satisfies the SINR constraint  \eqref{SINR_constr}. Some existing works propose different methods to address this issue. Here, we focus on deriving the optimal solution structure via the SCA method, not the implementation or numerical behavior of this method. Thus, we  only  assume a feasible initial point  $\zbf^{(0)}$ without discussing how to obtain it.

\subsection{The Optimal Multicast Beamforming Solution}\label{subsec:dual}
Since $\Pc_\SCA(\zbf)$ is convex (and Slater's condition holds), we  obtain its optimal solution from its Lagrange dual domain.
The Lagrangian for $\Pc_\SCA(\zbf)$ is given by
\begin{align}
\hspace*{-.5em}\Lc(\zbf,\wbf,\lambdabf)
=& \sum_{i=1}^G \|\wbf_i\|^2+\sum_{i=1}^G\sum_{k=1}^{K_i}\lambda_{ik} \bigg[ \gamma_{ik}\sum_{j\neq i}\left| {\wbf_j^H\hbf_{ik}} \right|^2 \nn \\
& \!\!\!  -2\mathfrak{Re}\left\{\wbf_i^H\hbf_{ik}\hbf_{ik}^H\zbf_i\right\}
+\left|\zbf_i^H\hbf_{ik}\right|^2\!+\sigma^2 \gamma_{ik}\bigg]
 \label{eq:Lagrangian}
\end{align}
where $\lambda_{ik}$ is the Lagrange multiplier associated with constraint \eqref{SCA_constr} for user $k$ in group $i$, and $\lambdabf \triangleq [\lambdabf_1^T,\ldots,\lambdabf_G^T]^T$ with $\lambdabf_i \triangleq [\lambda_{i1},\ldots,\lambda_{iK_i}]^T$. The Lagrange dual problem for $\Pc_\SCA(\zbf)$ is given by
\begin{align*} 
{\Dc_\SCA (\zbf):} \ \max_{\lambdabf} &\ g(\zbf,\lambdabf) \quad \st  \ \lambdabf \succcurlyeq {\bf 0}
\end{align*}
where
\begin{align} \label{dual_func}
g(\zbf,\lambdabf) \triangleq \min_{\wbf} \ & \Lc(\zbf,\wbf,\lambdabf).
\end{align}

Regrouping the different terms in \eqref{eq:Lagrangian}, the Lagrangian  can be rewritten as
\begin{align} \label{multi:newdual}
\Lc(\zbf,\wbf,\lambdabf) = &\sum_{i=1}^G\sum_{k=1}^{K_i}\lambda_{ik}\left(\sigma^2 \gamma_{ik}+\left|\zbf_i^H\hbf_{ik}\right|^2\right) \nn \\
& + \sum_{i=1}^G\wbf_i^H \left(\Ibf + \sum_{j\neq i} \sum_{k=1}^{K_j} \gamma_{jk}\lambda_{jk}\hbf_{jk}\hbf_{jk}^H\right)\wbf_i \nn \\
& - \sum_{i=1}^G 2\Re\left\{\zbf_i^H\left(\sum_{k=1}^{K_i} \lambda_{ik}\hbf_{ik}\hbf_{ik}^H\right)\wbf_i\right\}.
\end{align}
Define $\nubf_i \triangleq \left(\sum_{k=1}^{K_i} \lambda_{ik}\hbf_{ik}\hbf_{ik}^H\right)\zbf_i$, and
\begin{align}
\Rbf_{i^-}(\lambdabf) &\triangleq \Ibf + \sum_{j\neq i} \sum_{k=1}^{K_j} \lambda_{jk}\gamma_{jk}\hbf_{jk}\hbf_{jk}^H \label{Sigma}.
\end{align}
Then, the optimization problem   \eqref{dual_func} is equivalent to
\begin{align}\label{dual:g}
 \min_{\wbf} \sum_{i=1}^G \left( \wbf_i^H\Rbf_{i^-}(\lambdabf)\wbf_i
-2\Re\left\{\nubf_i^H\wbf_i\right\}\right).
\end{align}
The above optimization problem can be decomposed into subproblems with respect to (w.r.t.) each $\wbf_i$, $i \in \Gc$, as
\begin{align} \label{dual_w_i}
\min_{\wbf_i} \ \wbf_i^H\Rbf_{i^-}(\lambdabf)\wbf_i
-2\Re\{\nubf_i^H\wbf_i\},
\end{align}
and solved separately.
Since the optimization problem \eqref{dual_w_i} is convex, we can obtain its optimal solution in closed-form using KKT conditions \cite{Boyd:book}. The solution is given as follows.
\begin{proposition}\label{prop1}
The optimal solution for $\Pc_\SCA(\zbf)$ is given by
\begin{align}\label{w_o_SCA}
\wbf_i^\star(\zbf) &= \Rbf_{i^-}^{-1}(\lambdabf^\star)\left(\sum_{k=1}^{K_i}\alpha_{ik}^\star\hbf_{ik}\right), \ \ i \in \Gc 
\end{align}
where $\lambdabf^\star$ is the optimal dual solution for $\Dc_\SCA (\zbf)$, and $\alpha_{ik}^\star\triangleq \lambda_{ik}^\star\hbf_{ik}^H\zbf_i$, $k\in\Kc_i$,  $i \in \Gc$.
\end{proposition}
\IEEEproof
We first provide the complex gradients of two functions. Denote the real and imaginary part of vector $\xbf$ as $\xbf = \xbf_{\textrm{R}}+j\xbf_{\textrm{I}}$. For complex vector $\cbf$,  by the complex derivative operation  \cite{Kay:Book}, we have\begin{align}
\nabla{\xbf}\Re\{\cbf^H\xbf\}
&= \frac{1}{2}\left(\nabla{\xbf_{\textrm{R}}}\Re\{\cbf^H\xbf\}-\rm{j}\nabla{\xbf_{\textrm{I}}}\Re\{\cbf^H\xbf\}\right) \nn \\
&= \frac{1}{2}(\cbf_{\textrm{R}}-\rm{j}\cbf_{\textrm{I}})=\frac{1}{2}\cbf^*\label{appA:eq1}
\end{align}
where we note that $\Re\{\cbf^H\xbf\} = \cbf_{\textrm{R}}^T\xbf_{\textrm{R}}+\cbf_{\textrm{I}}^T\xbf_{\textrm{I}}$. Also, for Hermitian matrix $\Cbf $, we have
\begin{align}\label{prop1:eq2}
\nabla\xbf(\xbf^H\Cbf\xbf) = (\Cbf\xbf)^*.
\end{align}

The optimization problem \eqref{dual_w_i} is an unconstrained convex optimization problem. Denote the objective function in \eqref{dual_w_i} by $J(\zbf_i,\wbf_i)$ for given $\zbf_i$. Let $\lambdabf^\star$ be the optimal Lagrange multiplier vector for the dual problem $\Dc_\SCA(\zbf)$. By the KKT condition,  and from  \eqref{appA:eq1} and \eqref{prop1:eq2}, at the optimality of $\Pc_\SCA(\zbf)$, the gradient of $J(\zbf_i,\wbf_i)$ w.r.t. $\wbf_i$ satisfies
\begin{align}\label{prop1:eq3}
\nabla{\wbf_i}J(\zbf_i,\wbf_i)= \left(\Rbf_{i^-}(\lambdabf^\star)\wbf_i(\zbf)\right)^*-\nubf_i^*={\bf 0},
\end{align}
and we obtain
\begin{align}
\hspace*{-.5em}\wbf_i^\star(\zbf) &=\Rbf_{i^-}^{-1}(\lambdabf^\star)\nubf_i= \Rbf_{i^-}^{-1}(\lambdabf^\star)\left(\sum_{k=1}^{K_i} \lambda_{ik}^\star\hbf_{ik}\hbf_{ik}^H\right)\zbf_i \label{prop1:eq4}\\
&=\Rbf_{i^-}^{-1}(\lambdabf^\star)\left(\sum_{k=1}^{K_i}\alpha_{ik}^\star\hbf_{ik}\right)
\end{align}
where  $\alpha_{ik}^\star \triangleq \lambda_{ik}^\star\hbf_{ik}^H\zbf_i$, for $k\in\Kc_i$, $i\in\Gc$.
\endIEEEproof

 Examining the optimal  solution $\wbf_i^\star(\zbf)$ in Proposition~\ref{prop1}, we note that the dependency of $\wbf_i^\star(\zbf)$  on $\zbf$ is only through $\lambdabf^\star$ in  $\Rbf_{i^-}(\lambdabf^\star)$ and $\{\alpha_{ik}^\star\}$, both of which are functions of $\zbf$. This implies that, as the SCA method iteratively updates  $\zbf$, the  optimal solution $\wbf^\star(\zbf)$ for $\Pc_\SCA(\zbf)$ is updated accordingly, but only through $\lambdabf^\star$  and $\{\alpha_{ik}^\star\}$, while the structure of $\wbf^\star(\zbf)$ is unchanged. Thus, if $\zbf \to \wbf^o$, then the optimal solution for $\Pc_o$ is obtained.

Define $\Hbf_i \triangleq [\hbf_{i1},\ldots,\hbf_{iK_i}]$ as the channel matrix for group $i$,  and
\begin{align}\label{R}
\Rbf(\lambdabf) &\triangleq \Ibf + \sum_{i=1}^G \sum_{k=1}^{K_i} \lambda_{ik}\gamma_{ik}\hbf_{ik}\hbf_{ik}^H.
\end{align}
We state the main result in the following theorem.
 \begin{theorem}\label{thm1}
The optimal beamforming solution for the multi-group multicast beamforming QoS problem $\Pc_o$ is given by
\begin{align} \label{w_o2}
\wbf_i^o = \Rbf^{-1}(\lambdabf^o)\sum_{k=1}^{K_i}a_{ik}^o\hbf_{ik}=\Rbf^{-1}(\lambdabf^o)\Hbf_i\abf_i^o,  \  i\in \Gc
\end{align}
 where $\lambdabf^o$ is the optimal dual solution for $\Dc_\SCA (\wbf^o)$,   $a_{ik}^o \triangleq \lambda_{ik}^o\delta_{ik}(1+\gamma_{ik})$ with $\delta_{ik} \triangleq \hbf_{ik}^H\wbf_i^o$, $k\in\Kc_i$,  and $\abf_i^o \triangleq [a_{i1}^o,\ldots,a_{iK_i}^o]^T$,  $i \in \Gc$.
\end{theorem}
\IEEEproof
The SCA iterative procedure  described in Section~\ref{subsec:SCA}  is guaranteed to converge to a stationary point. This means that assuming  initial $\zbf^{(0)}$  chosen at the vicinity of the global optimal solution, the method will converge to the global optimal solution, \ie $\zbf\to \wbf^o$, and $\wbf_i^\star(\zbf)\to\wbf_i^o$.
Specifically,  from \eqref{prop1:eq4}, the optimal $\wbf_i^\star(\zbf)$ for $\Pc_\SCA(\zbf)$ in each iteration satisfies
\begin{align}\label{thm1:eq1}
\Rbf_{i^-}(\lambdabf^\star)\wbf_i^\star(\zbf) &=\sum_{k=1}^{K_i} \lambda_{ik}^\star\hbf_{ik}\left(\hbf_{ik}^H\zbf_i\right).
\end{align}
From \eqref{R}, we have $\Rbf_{i^-}(\lambdabf)=\Rbf(\lambdabf)-\sum_{k=1}^{K_i} \lambda_{ik}\gamma_{ik}\hbf_{ik}\hbf_{ik}^H$. Substituting this into \eqref{thm1:eq1}, we have
\begin{align}
\Rbf(\lambdabf^\star)\wbf_i^\star(\zbf)
&=  \sum_{k=1}^{K_i} \lambda_{ik}^\star\left(\hbf_{ik}^H\zbf_i+\gamma_{ik}\hbf_{ik}^H\wbf_i^\star(\zbf)\right)\hbf_{ik}. \label{thm1:eq2}
\end{align}

At the  convergence $\zbf\to \wbf^o$, we have $\hbf_{ik}^H\zbf_i\to \hbf_{ik}^H\wbf_i^o \triangleq \delta_{ik}$. Also, as $\zbf\to\wbf^o$, we have $\wbf_i^\star(\zbf) \to \wbf_i^o$, and the optimal $\lambdabf^\star$ for $\Dc_\SCA(\zbf)$ converges to $\lambdabf^o$ for $\Dc_\SCA(\wbf^o)$. Then, the  expression in \eqref{thm1:eq2} becomes
\begin{align}
\Rbf(\lambdabf^o)\wbf_i^o&=  \sum_{k=1}^{K_i} \lambda_{ik}^o\delta_{ik}(1+\gamma_{ik})\hbf_{ik},
\end{align}
and thus we have
\begin{align*}
\wbf_i^o&=\Rbf^{-1}(\lambdabf^o) \!\!  \sum_{k=1}^{K_i} \lambda_{ik}^o\delta_{ik}(1+\gamma_{ik})\hbf_{ik}
=\Rbf^{-1}(\lambdabf^o) \!\!  \sum_{k=1}^{K_i} a_{ik}^o\hbf_{ik}
\end{align*}
where $a_{ik}^o \triangleq \lambda_{ik}^o\delta_{ik}(1+\gamma_{ik})$, for $k\in \Kc_i$, $i\in\Gc$.
\endIEEEproof

Theorem~\ref{thm1} presents the   structure of the optimal multicast beamforming vector $\wbf_i^o$ for $\Pc_o$. The optimal solution $\wbf_i^o$   in \eqref{w_o2}  is expressed  in  semi-closed-form, where  $\lambdabf^o$ and $\abf_i^o$   need to be determined numerically. We point out that computing the optimal  $\lambdabf^o$ and $\abf_i^o$ is still challenging because $\Pc_o$ is NP-hard. This point will be revisited in Section~\ref{subsubsec:multi-unicast} when we compare multicast beamforming with unicast beamforming. In Section~\ref{sec:num}, we will provide numerical algorithms to compute $\lambdabf$ and $\abf_i$.

From  \eqref{thm1:eq1} and following  the proof of Theorem~\ref{thm1},
it is straightforward to show  the structure of the optimal beamforming vector $\wbf_i^o$  in an alternative form, given in the following corollary.
\begin{corollary}\label{cor1}
The optimal solution $\wbf_i^o$ in \eqref{w_o2} has the following equivalent alternative
form
\begin{align} \label{w_o}
\wbf_i^o &= \Rbf_{i^-}^{-1}(\lambdabf^o)\sum_{k=1}^{K_i}\!\alpha_{ik}^o\hbf_{ik}=\Rbf_{i^-}^{-1}(\lambdabf^o)\Hbf_i\alphabf_i^o, \  i\in \Gc
\end{align}
 where $\lambdabf^o_i$ is the same as  in Theorem~\ref{thm1}, and $\alphabf_i^o \triangleq [\alpha_{i1}^o,\ldots,\alpha_{iK}^o]^T$  with $\alpha_{ik}^o\triangleq\lambda_{ik}^o\delta_{ik}$, in which $\delta_{ik}$ is given in Theorem~\ref{thm1}.
\end{corollary}

Note that comparing the definition of $a_{ik}^o$ in \eqref{w_o2} with that of $\alpha_{ik}^o$ in \eqref{w_o}, we have the relation $\alpha_{ik}^o=a_{ik}^o/(1+\gamma_{ik})$, $k\in \Kc_i$, $i\in \Gc$.

The  value of the minimum power objective of $\Pc_o$ is given in the following corollary.
\begin{corollary}\label{cor2}
At the optimum of $\Pc_o$, the minimum power objective value is given by
\begin{align}\label{cor2_eq}
\sum_{i=1}^G \|\wbf_i^o\|^2&= \sigma^2\sum_{i=1}^G{\lambdabf^o_i}^T\gammabf_i = \sigma^2 {\lambdabf^o}^T\gammabf
\end{align}
where  $\gammabf\triangleq[\gammabf_1^T,\ldots,\gammabf_G^T]^T$ is the SINR\ target vector with   $\gammabf_i \triangleq [\gamma_{i1},\ldots,\gamma_{iK_i}]^T$, $i\in \Gc$, and $\lambdabf^o$ is given  in Theorem~\ref{thm1}.
\end{corollary}
\IEEEproof
See Appendix~\ref{app:cor2}.
\endIEEEproof

\vspace{0.5em}
\noindent{\bf Remark}  (\emph{Locally optimal multicast beamforming vector}): As mentioned in Section~\ref{subsec:SCA}, the SCA method  for the multicast beamforming problem $\Pc_o$ may converge to a local minimum. Following Proposition~\ref{prop1} and Theorem~\ref{thm1}, we have the  structure of any  locally optimal beamforming solution as follows.
\begin{corollary}\label{cor3}
Any locally optimal multicast beamforming solution $\wbf_i^\text{lo}$ for $\Pc_o$ has the following structure
\begin{align}\label{cor3:w_lo}
\wbf_i^\text{lo} &= \Rbf^{-1}(\lambdabf)\Hbf_i\abf_i, \ \ i \in \Gc
\end{align}
for some $\lambdabf$ and $\abf_i$, $i\in \Gc$.
\end{corollary}

Comparing \eqref{cor3:w_lo} with \eqref{w_o2}, we note that a locally  optimal multicast beamformer has a similar solution structure as the globally optimal one. The difference between the two lies in  the values of  $\lambdabf$ and $\abf_i$:  those  in \eqref{cor3:w_lo} obtained via the SCA method are suboptimal.

\subsection{Discussions on the Optimal Solution Structure}\label{subsec:discussion}
From  Theorem~\ref{thm1}, we have several important  observations on the structure of the optimal multicast beamforming solution, which are summarized below.

\subsubsection{Uplink-downlink duality interpretation}
Uplink-downlink duality has been established for the  downlink multi-user unicast  beamforming problem \cite{Rashid&Liu&Tassiulas:JSAC98,Visotsky&Madhow:VT99}, showing that the downlink beamforming problem  can be transformed into an equivalent uplink beamforming problem to solve. The structure of the optimal beamforming solution in \eqref{w_o2} indicates that there is a similar uplink-downlink duality interpretation for the downlink multi-group multicast beamforming problem as well. To see this, 
notice the following optimization problem
\begin{align} \label{dual_BF}
\max_{\wbf_i} \  \frac{\displaystyle \left|\wbf_i^H\left( \sum_{k=1}^{K_i}\lambda_{ik}^o\delta_{ik} \hbf_{ik}\right)\right|^2}{\displaystyle \sum_{j\neq i}\sum_{k=1}^{K_i}\lambda_{jk}^o\gamma_{jk}|\wbf_i^H\hbf_{jk}|^2+\wbf_i^H\wbf_i}
\end{align}
which can be rewritten as
\begin{align*}
\max_{\wbf_i} \frac{\wbf_i^H\tilde{\hbf}_i\tilde{\hbf}_i^H \wbf_i}{\wbf_i^H\Rbf_{i^-}(\lambdabf^o)\wbf_i}
\end{align*}
where $\tilde{\hbf}_i \triangleq \sum_{k=1}^{K_i}\lambda_{ik}^o\delta_{ik} \hbf_{ik}=\Hbf_i\alphabf_i^o$, with $\alphabf^o$  defined in Corollary~\ref{cor1}. The above problem is a generalized eigenvalue problem whose optimal solution is given by 
\begin{align*}
\wbf_i^s &=  \Rbf_{i^-}^{-1}(\lambdabf^o)\tilde{\hbf}_i=\Rbf_{i^-}^{-1}(\lambdabf^o)\Hbf_i\alphabf_i^o =\Rbf^{-1}(\lambdabf^o)\Hbf_i\abf_i^o
\end{align*}
where the last equation is by Corollary~\ref{cor1}, and the solution is identical to \eqref{w_o2}. Thus, the optimal beamforming vector $\wbf_i^o$ in \eqref{w_o2} is  the solution to the optimization problem~\eqref{dual_BF}.

The optimization problem~\eqref{dual_BF} can be interpreted as an uplink  receive beamforming problem for SINR\ maximization: For an uplink system with multiple receiver antennas, consider the dual uplink channel  $\hbf_{ik}$, transmit power $P_{ik}=\sigma^2\lambda_{ik}^o\gamma_{ik}$ for  user $k$ in group $i$, and the receiver noise covariance $\sigma^2\Ibf$. Then, the problem \eqref{dual_BF} is equivalent to the following  problem
\begin{align} \label{dual_BF2}
\max_{\wbf_i} \  \frac{\displaystyle \left|\wbf_i^H\left( \sum_{k=1}^{K_i}\tilde{\delta}_{ik}\sqrt{P_{ik}} \hbf_{ik}\right)\right|^2}{\displaystyle \sum_{j\neq i}\sum_{k=1}^{K_j}P_{jk}|\wbf_i^H\hbf_{jk}|^2+\sigma^2\wbf_i^H\wbf_i}
\end{align}
where $\tilde{\delta}_{ik}\triangleq \delta_{ik}\sqrt{\lambda_{ik}^o/\gamma_{ik}}$. The  problem \eqref{dual_BF2} can be interpreted as the optimal uplink beamforming to maximize the receiver SINR at a \emph{group-channel direction}. This group-channel direction is specified by the weighted sum of channels in group $i$, defined by  $\sum_{k=1}^{K_i}\tilde{\delta}_{ik} \sqrt{P_{ik}}\hbf_{ik}$, where $\tilde{\delta}_{ik}$ is the weight for each user in the group.

Note that in the uplink beamforming problem \eqref{dual_BF2},
  $\{\tilde{\delta}_{ik}\}$ and $\{P_{ik}\}$ are given. These need to be obtained for $\wbf_i^o$ in $\Pc_o$ (\ie the optimal $\lambdabf^o$ and $\{\delta_{ik}\}$). These parameters specify the group-channel direction and need to be determined via other methods. This is the difference  between  multicast beamforming and unicast beamforming on the uplink-downlink duality. For the unicast beamforming, the related parameter in the optimal beamforming vector can be determined via optimizing the dual uplink power allocation to minimize the sum-power \cite{Rashid&Liu&Tassiulas:JSAC98,Visotsky&Madhow:VT99}.

\subsubsection{Weighted MMSE beamforming structure}
 For multi-user uplink transmissions, it is known that the optimal receive beamforming vector for SINR\ maximization is the MMSE filter. Following the uplink-downlink duality interpretation  for multicast beamforming, we see that this is indeed  the structure of $\wbf_i^o$  given in \eqref{w_o2}.  More precisely, the solution structure indicates that the optimal multicast beamforming vector $\wbf_i^o$ is a weighted MMSE filter. The  optimal $\wbf_i^o$ contains two terms:
\begin{itemize}
\item A weighted sum of channel vectors of the intended user group $i$: $\sum_{k=1}^{K_i}a_{ik}^o\hbf_{ik}\triangleq \hat{\hbf}_i$. The resulting $\hat{\hbf}_i$ is the multicast  \emph{group-channel direction}.\footnote{The group-channel direction $\hat{\hbf}_i$ can be defined up to a scaling factor: $\hat{\hbf}_i=c\sum_{k=1}^{K_i}a_{ik}^o\hbf_{ik}$, for $c$ being a scaler.} Weight $a_{ik}^o$ determines the relative significance of   user $k$'s channel  $\hbf_{ik}$ in this group-channel direction.
\item Matrix $\Rbf(\lambdabf)$ is the  (normalized)  noise plus weighted channel covariance (of all groups)  matrix   (and likewise, $\Rbf_{i^-}(\lambdabf)$ is  the (normalized) noise plus weighted interference covariance  matrix for group $i$), where $\lambda_{ik}\gamma_{ik}$ is the  weight of each user channel  $\hbf_{ik}$ relative to others.\footnote{For convenience, here we refer to $\hbf_{ik}\hbf_{ik}^H$ as the channel covariance matrix, considering $\hbf_{ik}$ is given  deterministic.}
\end{itemize}

In the special case of a single user per group ($K=1$), the system reduces to the traditional downlink unicast multi-user  beamforming problem.  For notation simplicity, we remove subscript $k$ in the notations to represent the unicast case, and the expression of beamforming solution $\wbf_i^o $ in \eqref{w_o2} reduces to
\begin{align}\label{w_o_unicast}
\wbf_i^o = a_i^o\left(\Ibf + \sum_{i=1}^G \lambda_{i}^o\gamma_{i}\hbf_{i}\hbf_{i}^H\right)^{-1}\!\!\hbf_i,
\end{align}
 which is exactly the classical downlink multi-user unicast beamforming solution \cite{Schubert&Boche:TVT04,Bjornson&Bengtsson&Ottersten:SPM14}.

\subsubsection{Multicast versus unicast}\label{subsubsec:multi-unicast}
Comparing the optimal beamforming structures in  \eqref{w_o2} and  \eqref{w_o_unicast}, we can view the  optimal multicast beamforming as the generalized version of  the optimal  unicast beamforming. It is a weighted MMSE filter with a similar covariance matrix structure, except that the signal direction is now a multicast group-channel direction instead of the unicast individual user channel direction.\footnote{Alternatively, we may also interpret the optimal structure as the weighted optimal unicast beamforming vectors, with the weight giving different emphasis on each user's beamforming vector based on its channel condition. }

While structurally similar,  there is a key difference between the optimal $\wbf_i^o$  in multicast and in unicast. For the unicast beamforming QoS problem, the SINR target constraint for each user is attained with equality at  optimality. This allows the optimal $a_i^o$ in \eqref{w_o_unicast} to be solved easily for the optimal $\wbf_i^o$. In contrast, for  multicast beamforming, the SINR constraints  will not be all attained with equality in general. This  uncertainty adds difficulty in determining the optimal weight vector $\abf_i^o$ for the optimal $\wbf_i^o$, which reflects the NP-hard nature of the multicast beamforming problem $\Pc_o$. As a result, we obtain the structure of the optimal solution   $\wbf_i^o$ in  \eqref{w_o2}, while the optimal weights $\{\abf_i^o\}$ and $\lambdabf^o$ are still challenging to determine.  In Section~\ref{sec:num}, we propose numerical algorithms to compute them.

\subsubsection{Inherent low-dimensional structure}
One main issue of existing numerical methods to compute a feasible multicast beamforming  solution is their computational complexity, which has a high order of growth w.r.t. the number of antennas $N$, making them unrealistic for practical implementation in massive MIMO systems   with $N\gg 1$. Some recent works \cite{LeNam&etal:SPL14,Christopoulos&etal:SPAWC15,Sadeghi&etal:TWC17} have proposed reduced complexity algorithms to reduce the scaling order of  complexity   w.r.t. $N$.

An  important observation of the optimal multicast beamforming vector $\wbf_i^o$  in  \eqref{w_o2} is that it has an inherent low-dimensional structure for computation. As mentioned earlier, the solution is based on a weighted sum of channel vectors in the group. Instead of directly optimizing  $\wbf_i$ of  $N$-dimension, the problem is equivalent to optimizing the weight vector $\abf_i$ of $K_i$-dimension (details are given in    Section~\ref{subsec:weight_a}). For systems with $N\gg K_i$,  this means a significant reduction of the  complexity in computing  the beamforming solution. This low-dimensional structure in the solution brings an immediate benefit to the multicast beamforming design in massive MIMO systems, where typically we expect the number of antennas is much more than the size of each multicast user group ($N\gg K_i$). Optimizing  weight vector $\abf_i$, instead of $\wbf_i$ directly, reduces the size of optimization variables to $K_i$. As a result, the computational complexity will no longer grow with $N$. This  leads to substantial  computational saving, which lifts the computational barrier for designing multicast beamforming  in massive MIMO\ systems.

In general, depending on the values of $N$ and $K_i$, we can choose to directly solve $\wbf_i$ or  weight vector $\abf_i$, whichever has a lower dimension, to minimize the computational complexity in finding the beamforming solution.
This applies to both the traditional multi-antenna systems and massive MIMO systems. Furthermore, note that  $\Hbf_i\abf_i$ in the optimal $\wbf_i^o$  in  \eqref{w_o2}  is a linear combination of channel vectors. This  suggests that we can further reduce the size of the weight optimization problem  by only considering the dimension of the channel space spanned by $\Hbf_i$. Assume that the $N\times K_i$ channel matrix $\Hbf_i$ for group $i$ has rank $r_i\le \min(N,K_i)$. Let $\Ubf_i\triangleq[\ubf_{i1},\ldots,\ubf_{ir_i}]$ be the $N\times r_i$ matrix containing the orthonormal vectors that span the column space of $\Hbf_i$.\footnote{The SVD of $\Hbf_i$ is $\Hbf_i = [\Ubf_i \ \ \Ubf_i^\bot] \Lambdabf\Vbf$, with $\Ubf_i$ consisting of the left singular vectors corresponding to the first $r_i$ non-zero singular values in $\Lambdabf$.} Then, we can express $\Hbf_i\abf_i$ as
\begin{align}
\Hbf_i\abf_i = \Ubf_i\bbf_i
\end{align}
for some $r_i\times 1$  vector $\bbf_i$. Thus, the weight optimization problem w.r.t $\abf_i$ can be further transformed into a size-reduced weight optimization problem w.r.t.  $\bbf_i$ of $r_i$-dimension. Methods used to solve $\{\abf_i\}$, as described in Section~\ref{subsec:weight_a} can be similarly applied to solve $\{\bbf_i\}$.

\section{Numerical Algorithms and Analysis}\label{sec:num}
The  optimal multicast beamforming solution $\wbf_i^o$ in \eqref{w_o2} contains parameters  $\lambdabf^o$ and $\abf_i^o$   that need to be computed numerically. As discussed in Section~\ref{subsec:discussion}, obtaining  the optimal  $\lambdabf^o$ and  $\abf_i^o$ is challenging, due to the NP-hard nature of  $\Pc_o$. In this section, we develop numerical algorithms to compute  $\lambdabf$ and  $\abf_i$.

\subsection{Algorithm for Lagrange Multiplier  $\lambdabf$}\label{subsec:lambda_alg}

 Define $\Dbf_{\gammabf_i}\triangleq\diag(\gammabf_i)$, $\Dbf_{\lambdabf_i}\triangleq\diag(\lambdabf_i)$,  and $\deltabf_i \triangleq [\delta_{i1},\ldots,\delta_{iK_i}]^T$, $i\in\Gc$. The definition of $\abf_i$ is given in Theorem~\ref{thm1}. We express it in a compact matrix form as $\abf_i = \Dbf_{\lambdabf_i}(\Ibf+\Dbf_{\gammabf_i})\deltabf_i$. By Theorem~\ref{thm1}, at  optimality, we have
\begin{align}
\delta_{ik} = \hbf_{ik}^H\wbf_i^o
= \hbf_{ik}^H \Rbf^{-1}(\lambdabf)\Hbf_i\Dbf_{\lambdabf_i}(\Ibf+\Dbf_{\gammabf_i})\deltabf_i,
\end{align}
for $k\in K_i, i\in \Gc$.
It follows that
\begin{align}
\deltabf_i &= \Hbf_i^H\Rbf^{-1}(\lambdabf)\Hbf_i\Dbf_{\lambdabf_i}(\Ibf+\Dbf_{\gammabf_i})\deltabf_i, \ i\in \Gc,
\end{align}
which is equivalent to  \begin{align}\label{lambda_opt1}
\left(\Hbf_i^H\Rbf^{-1}(\lambdabf)\Hbf_i\Dbf_{\lambdabf_i}(\Ibf+\Dbf_{\gammabf_i})-\Ibf\right)\deltabf_i = {\bf 0}.
\end{align}

At the optimality, the optimal $\lambdabf^o$ should satisfy \eqref{lambda_opt1}.
However, since $\deltabf_i$ is unknown,  directly solving \eqref{lambda_opt1} is difficult.\footnote{For \eqref{lambda_opt1} to hold, $\deltabf_i$ should be in the null space of matrix  $\Hbf_i^H\Rbf^{-1}(\lambdabf)\Hbf_i\Dbf_{\lambdabf_i}(\Ibf+\Dbf_{\gammabf_i})-\Ibf$. However, with unknown $\deltabf_i$, it is difficult to use this condition to derive $\lambdabf^o$.} Instead, we propose a suboptimal algorithm to compute $\lambdabf$ below, which we later show to be asymptotically optimal as $N\to \infty$.

A sufficient condition to satisfy  \eqref{lambda_opt1} is  the following
\begin{align}\label{lambda_mtx_eqn}
\Hbf_i^H\Rbf^{-1}(\lambdabf)\Hbf_i\Dbf_{\lambdabf_i}(\Ibf+\Dbf_{\gammabf_i})=\Ibf, \quad i \in \Gc.
\end{align}
which is equivalent to, for $ i \in \Gc$,
\begin{align}\label{lambda_eqn}
\begin{cases}\lambda_{ik} (1+\gamma_{ik})\hbf_{ik}^H\Rbf^{-1}(\lambdabf)\hbf_{ik} = 1, &  k \in \Kc_i, \\
\lambda_{ik} (1+\gamma_{ik})\hbf_{ik}^H\Rbf^{-1}(\lambdabf)\hbf_{il} = 0, & l \neq k, l\in \Kc_i.
\end{cases}
\end{align}
However, the above conditions may not be satisfied for all $\lambda_{ik}$, since there are typically  more equations than variables to solve. In the following, we propose to obtain $\lambdabf$ by only solving  the first equation in \eqref{lambda_eqn},  \ie
\begin{align}\label{lambda_eqn1}
\lambda_{ik} (1+\gamma_{ik})\hbf_{ik}^H\Rbf^{-1}(\lambdabf)\hbf_{ik} = 1, \ \  k \in \Kc_i, i\in\Gc.
\end{align}
The solution $\lambdabf$ to the above fixed-point equations can be obtained using the fixed-point iterative method as follows:
\begin{enumerate}
\item Initialize $\lambdabf^{(0)}$; Set $l=1$.
\item Compute $\lambda_{ik}^{(l)}$: for  each $k \in \Kc_i, i\in\Gc$,
\begin{align}\label{lambda_fixpoint}
\lambda_{ik}^{(l)} &= \frac{1}{(1+\gamma_{ik})\hbf_{ik}^H\Rbf^{-1}(\lambdabf^{(l-1)})\hbf_{ik} }.
\end{align}
\item Set $l=l+1$; Repeat Steps 2-3 until convergence.
\end{enumerate}


\subsection{Asymptotic Analysis of $\lambdabf$}\label{subsubsec:AS}

Our   solution for  $\lambdabf$ by the proposed algorithm has the following asymptotic property as the number of antennas $N$ grows.
\begin{proposition} \label{prop2}
Assume channel vectors $\hbf_{ik}$'s are independent, and the elements $h_{ik,n}$'s in  $\hbf_{ik}$ are i.i.d. with $E(h_{ik,n})=0$ and $E(|h_{ik,n}|)<\infty$. As  $N\to \infty$, the solution $\lambdabf$ of \eqref{lambda_eqn1} satisfies \eqref{lambda_opt1} almost surely, and thus  is asymptotically optimal.
\end{proposition}
\IEEEproof See Appendix~\ref{app:prop2}.\endIEEEproof

Note that the channel conditions in Proposition~\ref{prop2}  hold for  commonly used fading channel models, such as Rayleigh fading, where channels are  zero-mean Gaussian distributed.

The above results indicate that our algorithm to compute  $\lambdabf$ is  particularly efficient and effective  in massive MIMO systems with large $N$. The iterative procedure to compute $\lambdabf$ is simple with low computational complexity. There are $\sum_{i=1}^GK_i$ elements in $\lambdabf$ to be computed, which does not grow with  $N$.   At the same time, the asymptotic result in Proposition~\ref{prop2} indicates that   $\lambdabf$  computed by our proposed algorithm would be close to the optimal $\lambdabf^o$ for large $N$. We will see from the simulation  that our algorithm provides a near-optimal performance for a moderate value of $N$.

We further provide the asymptotic expression of $\lambdabf$ as $N \to \infty$. Let $\hbf_{ik}=\sqrt{\beta_{ik}}\gbf_{ik}$, where $\gbf_{ik}\sim \Cc\Nc({\bf 0},\Ibf)$, and $\beta_{ik}$ represents the large-scale channel variation.

\begin{proposition} \label{prop3}
Assume that channel vectors $\hbf_{ik}$'s are independent. As $N\to \infty$, the solution $\lambdabf$ for  \eqref{lambda_eqn1} is given by
\begin{align}\label{prop3:eqn0}
\hspace*{-.5em}\lambda_{ik}\beta_{ik} = \frac{1}{\displaystyle N-\mathop{\sum\sum}_{jl\neq ik}\gamma_{jl}}+o\left(\frac{1}{N^2}\right), \  k\in \Kc_i,i\in \Gc.
\end{align}
\end{proposition}
\IEEEproof See Appendix~\ref{app:prop3}.
\endIEEEproof

Proposition~\ref{prop3} shows the asymptotic behavior of $\lambdabf$ produced by our proposed algorithm.  For large $N$, $\lambda_{ik}\beta_{ik}$ can be approximated using the first term in \eqref{prop3:eqn0}, which greatly simplifies the computation of  $\lambdabf$, especially in massive MIMO\ systems. We also have the following important observations:
\subsubsection{Asymptotic $\Rbf(\lambdabf)$}
 For large $N$, the difference among $\lambda_{ik}\beta_{ik}$'s diminishes, and all $\lambda_{ik}\beta_{ik}$'s converge to nearly the same value. In the special case when the target SINRs for all users are equal,  $\gamma_{ik}=\gamma$, $\forall i,k$, all $\lambda_{ik}\beta_{ik}$'s converge to  the same value given by
\begin{align}\label{prop3:gamma}
\lambda_{ik}\beta_{ik} = \frac{1}{N- (K_\tot-1)\gamma}+o\left(\frac{1}{N^2}\right)
\end{align}
where recall that $K_\tot$ is the number of all users. Note from  $\Rbf(\lambdabf)$ in \eqref{R} that, $\lambda_{ik}$ is the weight for each user channel covariance term in $\Rbf(\lambdabf)$. The above indicates that, asymptotically, $\lambda_{ik}$ acts to \emph{normalize} the channel variance $\beta_{ik}$ for $\hbf_{ik}$ in $\Rbf(\lambdabf)$, which can be written as $\Rbf(\lambdabf) = \Ibf + \sum_{i=1}^G \sum_{k=1}^{K_i} (\lambda_{ik}\beta_{ik})\gamma_{ik}\gbf_{ik}\gbf_{ik}^H$. As a result, each user contribution in $\Rbf(\lambdabf)$  is equalized and  weighted only based on $\gamma_{ik}$'s (weighted equally when all  $\gamma_{ik}$'s are the same). This leads to a much-simplified  approximation in computing $\Rbf(\lambdabf)$ and thus  the optimal  $\wbf_i^o$ in \eqref{w_o2} in practice when $N$ is large. For example, in the case considered in \eqref{prop3:gamma}, we have
\begin{align}\label{prop3:R}
\Rbf(\lambdabf) \approx \Ibf + \frac{1}{\frac{N}{\gamma}-(K_\tot-1)} \sum_{i=1}^G
\sum_{k=1}^{K_i} \gbf_{ik}\gbf_{ik}^H,
\end{align}
provided that $N> (K_\tot-1)\gamma$.

\subsubsection{Slow diminishing rate of interference} We also  draw the following cautious observation. It is known  that, for transmit beamforming, interference at each user diminishes  as $N\to \infty$. The asymptotic   beamformer design and analysis in massive MIMO systems may be simplified by removing the interference, which is  considered for multi-group multicast beamforming  \cite{Xiang&Tao&Wang:IJSAC14,Sadeghi&Yuen:Globecom15}. This diminishing interference is similarly manifested in $\Rbf(\lambdabf)$ in \eqref{prop3:R}, where, as $N\to \infty$, $\Rbf(\lambdabf)$ converges to $\Ibf$, and $\wbf^o$ reduces to the weighted MRT beamforming.   However, the interference  may diminish slowly as $N$ increases,  and it  requires very large $N$ in practice to reflect the  asymptotic  behavior accurately.\footnote{This slow converging behavior is observed  in \cite{Sadeghi&Yuen:Globecom15,Yu&Dong:ICASSP18}, where the asymptotic beamformer (ignoring interference) performs poorly in a wide  range of $N$ values.} To see this, note that the total interference term in $\Rbf(\lambdabf)$ in \eqref{prop3:R} is reduced by  approximately a factor of $(N/\gamma-K_\tot)$, where  both $\gamma$ and $K_\tot$ affect the reduction rate over $N$.
For example, for  $G=3$, $K_i = 5$, $\forall i$, and $\gamma = 10$ dB, we have $N/\gamma=N/10$. For $N=256$,   $N/\gamma-K_\tot\approx 10$, and  the interference term in $\Rbf(\lambdabf)$ is  still non-negligible.  For $\gamma=20$ dB, it would require $N$ to be more than $2000$ for $N/\gamma-K_\tot\approx 10$. The above discussion shows that for the practical value of $N$ used in large-scale antenna systems,  the interference may still be substantial in the received SINR, and we need to consider it  in obtaining the optimal $\wbf_i^o$ in $\eqref{w_o2}$.

\subsection{Algorithms for Weight Vector  $\{\abf_i\}$} \label{subsec:weight_a}
Using the expression of the optimal $\wbf_i^o$  in \eqref{w_o2} and $\lambdabf$ computed by our algorithm in  Section~\ref{subsec:lambda_alg}, the multi-group multicast beamforming problem $\Pc_o$ w.r.t. $\wbf$ can be transformed into a weight  optimization problem w.r.t. $\abf\triangleq [\abf_1^H,\ldots,\abf_G^H]^H$ as follows \footnote{By Theorem~\ref{thm1}, $a_{ik} = \lambda_{ik}\delta_{ik}(1+\gamma_{ik})$. Thus, alternatively, we can  obtain $\{\deltabf_i\}$  for given $\{\lambdabf_i\}$ and$\{\gammabf_i\}$ by formulating a problem w.r.t. $\{\deltabf_i\}$ similar to $\Pc_1$. There is no difference in the two approaches, and we choose to directly obtain $\abf$ for simplicity.}
\begin{align}\label{opt_a}
\Pc_1: \ \min_{\abf} \ &  \sum_{i=1}^G\|\Rbf^{-1}(\lambdabf)\Hbf_i\abf_i\|^2
\nn \\
\st \ &  \displaystyle  \frac{|\abf_i^H\Hbf_i^H\Rbf^{-1}(\lambdabf)\hbf_{ik}|^2}{\displaystyle \sum_{j\neq i}  |\abf_j^H\Hbf_j^H\Rbf^{-1}(\lambdabf)\hbf_{ik}|^2+\sigma^2}\ge \gamma_{ik}, \nn \\[-1em]
&\hspace{1.5in}\ k\in \Kc_i,i\in \Gc.
\end{align}

The  optimization problem $\Pc_1$ is still NP-hard, since the form of constraints is similar to that in the original  problem $\Pc_o$. However, the key difference here is that the beamforming vector $\wbf$ in $\Pc_o$ is of size   $GN$, and in contrast, the weight vector $\abf$ for the weight optimization problem $\Pc_1$ is of size  $\sum_{i=1}^GK_i$, which no longer depends on $N$.  This is especially appealing to massive MIMO systems with $N\gg K_i$, $i\in \Gc$, because of a significant  computational saving by solving the much smaller problem $\Pc_1$ instead of $\Pc_o$.

As mentioned earlier, existing prevailing numerical algorithms for this family of problems have high computational complexity  for a large problem size (\eg SCA and SDR). This makes them impractical to directly compute  multicast beamforming solutions for large $N$.  Using the optimal beamforming structure in Theorem~\ref{thm1}, the  numerical computation   of the solution via $\Pc_1$ is no longer affected by $N$; It can be   done efficiently with low-complexity. Moreover, as the performance of some approaches may deteriorate  as the problem size grows, keeping the problem size small will maintain the quality of the computed solution.

In the following, we apply two approaches to compute the weight vector $\abf$ for $\Pc_1$.
\subsubsection{The SDR method}\label{subsec:weight_a_SDR}
Define $\Gbf_i \triangleq \Rbf^{-1}(\lambdabf)\Hbf_i$, and  $\fbf_{jik}\triangleq \Gbf_j^H\hbf_{ik}$, $k\in\Kc_i$, $i,j\in \Gc$. Define $\Xbf_i \triangleq \abf_i\abf_i^H$, $i\in \Gc$. Dropping the rank-one constraint on $\Xbf_i$, $\Pc_1$ is relaxed to the following SDP problem
\begin{align}
\Pc_{1\text{\tiny SDR}}:  & \min_{\{\Xbf_i\}}   \sum_{i=1}^G\tr(\Gbf_i^H\Gbf_i\Xbf_i)  \nn \\
\st  &  \left(\!\frac{1}{\gamma_{ik}}\!+1\!\right)\!\tr(\fbf_{iik}\fbf_{iik}^H\Xbf_i)\!-\!\!\sum_{j=1}^G\!\tr(\fbf_{jik}\fbf_{jik}^H\Xbf_j )\ge \sigma^2  \nn \\
&\Xbf_i \succcurlyeq  0,\hspace{1.4in}\ k\in \Kc_i,i\in \Gc. \nn
\end{align}

Standard SDP solvers can be used to solve $\Pc_{1\text{\tiny SDR}}$ to obtain the optimal  $\{\Xbf_i^o\}$. Finally,  $\{\abf_i^\text{\tiny SDR}\}$ can be extracted from $\{\Xbf_i^o\}$ by using the Gaussian randomization methods \cite{Sidiropoulosetal:TSP06}. Rank reduction based techniques \cite{HuangPalomar:TSP10} can also be applied to obtain $\{\abf_i^\text{\tiny SDR}\}$, depending on the number of constraints.

As mentioned above, a major benefit of adopting the
SDR method to solve $\Pc_1$, as compared to directly solving $\Pc_o$ by
the SDR, is the significantly smaller problem size  $\Pc_1$.
Specifically,   the complexity of solving $\Pc_1$ via SDP \cite{Bobo&Vandenberghe&Boyd:LAA98} is $\Oc((\sum_{i=1}^GK_i^2)^3)$, while the complexity of directly solving $\Pc_o$ via SDP is $\Oc((GN^2)^3)$.

\subsubsection{The SCA method}\label{subsec:weight_a_SCA}
We can apply the SCA method to iteratively solve $\Pc_1$ for $\abf$. Similar to $\Pc_\SCA(\zbf)$ in Section~\ref{subsec:SCA}, using  $K_i\times 1$ auxiliary vector $\vbf_i$, $i\in \Gc$, and applying the convex approximation to constraint \eqref{opt_a} in $\Pc_1$, we have the following convex optimization problem for any given $\vbf\triangleq [\vbf_1^H,\ldots,\vbf_G^H]^H$
\begin{align}
& \Pc_{1\SCA}(\!\vbf\!):  \min_{\{\abf_i\}}   \sum_{i=1}^G \|\Gbf_i\abf_i\|^2 \nn \\
&\st  \sum_{j=1}^G  |\abf_j^H\fbf_{jik}|^2 -2\left(\!\frac{1}{\gamma_{ik}}\!+1\!\right)\
\Re\{\abf_i^H\fbf_{iik}\fbf_{iik}^H\vbf_i\} \nn \\
& \quad\quad +\left(\!\frac{1}{\gamma_{ik}}\!+1\!\right)|\vbf_i^H\fbf_{iik}|^2 \le-\sigma^2, \ k\in \Kc_i, i \in \Gc.
\end{align}

 To obtain $\abf$ for  $\Pc_1$, iteratively solve $\Pc_{1\SCA}(\vbf)$ and update $\vbf$  with the optimal solution $\abf_i^\star(\vbf)$ for $\Pc_{1\SCA}(\vbf)$ until convergence. The steps are similar to those given  in Section~\ref{subsec:SCA}, and the convergence is standard. Note that solving  $\Pc_{1\SCA}(\vbf)$   in each SCA iteration using the typical interior-point method \cite{Boyd:book} has a  complexity of $\Oc((\sum_{i=1}^GK_i)^3)$, as opposed to $\Oc((GN)^3)$ for  $\Pc_{\SCA}(\zbf)$ in each iteration to directly solve $\Pc_o$.

\emph{Initialization}: In the SCA method, the initial $\vbf^{(0)}$ should be feasible to $\Pc_{1\SCA}(\vbf)$. To ensure this and expedite the convergence, we use the solution $\{\abf_i^{\text{\tiny SDR}}\}$ by the SDR method to set $\vbf_i^{(0)}=\abf_i^{\text{\tiny SDR}}$, $i\in\Gc$. The solution  $\{\abf_i^{\text{\tiny SDR}}\}$ provides a good initial point close to the optimum; it will fasten the convergence of the SCA method, and   increase the chance to converge to the global optimum (instead of a local optimum). In particular, since the problem size of $\Pc_{1\text{\tiny SDR}}$ is  small, computing  $\{\abf_i^{\text{\tiny SDR}}\}$ is fast  even for large $N$, adding minimal computational burden. 
This is verified by simulation in Section~\ref{sec:sim}.

\vspace*{0.5em}
\noindent{\bf Remark}: Finally, we point out that in considering the  above two  prevailing methods to solve $\Pc_1$ for  $\abf$, we emphasize  the computational and performance benefits of transforming $\Pc_o$ into the  weight optimization problem  $\Pc_1$ of a much smaller size. 
Other methods can be used to solve  $\Pc_1$ as well. In particular,    methods developed to solve $\Pc_o$  can be applied to solve $\Pc_1$, since $\Pc_o$ and $\Pc_1$ are structurally the same, and the  benefits   mentioned above also carry to these possible methods. For example,  the ADMM method \cite{Boyd&etal:2011ADMM} can be used to solve $\Pc_{\SCA}(\zbf)$ (or $\Pc_{1\SCA}(\vbf)$) in each SCA iteration; it can decouple the problem into per user  group subproblems with a reduced number of variables for faster computation. An ADMM-based algorithm has been recently proposed  \cite{Chen&Tao:TCOM17} for  the multicast beamforming problem (\eg $\Pc_o$). There may be other first-order approximation methods  to solve $\Pc_{\SCA}(\zbf)$. As mentioned above, these methods can also be  adopted to solve $\Pc_1$ to  reduce the computational complexity further.

\section{Multicast Beamforming for the MMF Problem} \label{sec:MMF}
In this section, we consider the  weighted MMF problem $\Sc_o$ for multi-group multicast beamforming and discuss how our results obtained for the QoS problem $\Pc_o$  can be extended to solve  $\Sc_o$.    We first transform  $\Sc_o$  into the following equivalent problem
\begin{align}
\Sc_1: \  \max_{\wbf,t}  & \ \ t \nn \\
\st & \ \ \SINR_{ik} \ge t\gamma_{ik}, \ k\in \Kc_i,\ i \in \Gc \\
& \ \ \sum_{i=1}^G \|\wbf_i\|^2 \le P. \nn
\end{align}

It has been shown that the QoS problem  $\Pc_o$ and the MMF problem $\Sc_1$ are inverse problems \cite{Karipidisetal:TSP08}. Specifically, for given SINR target vector $\gammabf$ and power budget $P$, explicitly parameterize the  problem $\Sc_1$ as $\Sc_1(\gammabf,P)$, with the optimal objective value as $t^o=\Sc_1(\gammabf,P)$. Also, parameterize $\Pc_o$ as $\Pc_o(\gammabf)$, with the minimum power as $P=\Pc_o(\gammabf)$. Then, the inverse relation of  problems $\Pc_o$ and $\Sc_1$ is described below
\begin{align}
t^o &= \Sc_1(\gammabf,\Pc_o(t^o\gammabf)), \label{eqn:inv_PtoS}\\
P &= \Pc_o(\Sc_1(\gammabf,P)\gammabf). \label{eqn:inv_StoP}
\end{align}

This inverse relation means that, if the  solution for $\Pc_o$ can be obtained, we can find the solution for $\Sc_1$ via iteratively solving $\Pc_o$ along with  a bi-section search over $t$ until the transmit power is equal to $P$. This procedure immediately implies that the optimal  beamforming vector for the MMF problem $\Sc_1$ has a similar structure as in \eqref{w_o2} for the QoS problem: A weighted MMSE filter with the group-channel direction formed by a weighted sum of channels in the group. Following this, as well as the relations in    \eqref{eqn:inv_PtoS} and \eqref{eqn:inv_StoP}, we have the optimal beamforming vector for  $\Sc_1$ given below.
\begin{theorem}\label{thm2}
The optimal beamforming solution for the MMF multi-group multicast beamforming problem $\Sc_o$ is given by
\begin{align} \label{w_o2_MMF}
\wbf_{\MMF,i}^o = \widetilde{\Rbf}^{-1}(\lambdabf_\text{\tiny QoS}^o)\Hbf_i\tilde{\abf}_i^o,  \  i\in \Gc
\end{align}
where  $\lambdabf_\text{\tiny QoS}^o$ is obtained from the optimal beamforming vector $\wbf_{\QoS,i}^o$ in \eqref{w_o2}   for the QoS problem $\Pc_o(t^o\gammabf)$,
\begin{align}\label{R_tilde}
 \widetilde{\Rbf}(\lambdabf_\text{\tiny QoS}^o)\defeq \Ibf + \frac{P}{\sigma^2} \sum_{i=1}^G \sum_{k=1}^{K_i}  \frac{\lambda_{\text{\tiny QoS},ik}^o\gamma_{ik}}{\lambdabf_\text{\tiny QoS}^{oT}\gammabf}\hbf_{ik}\hbf_{ik}^H,
\end{align}
and
 $\tilde{\abf}_i^o = [\tilde{a}_{i1}^o,\ldots,\tilde{a}_{iK_i}^o]^T$ with
 \begin{align}\label{a_MMF}
\tilde{a}_{ik}^o \triangleq \lambda_{\text{\tiny QoS},ik}^o\delta_{ik}\left(1+\frac{P\gamma_{ik}}{\sigma^2\lambdabf_\text{\tiny QoS}^{oT}\gammabf}\right)
\end{align} in which $\delta_{ik} = \hbf_{ik}^H\wbf^o_{\text{\tiny QoS},i}$, $k\in\Kc_i,i \in \Gc$.

 The optimal objective value $t^o$ of problem $\Sc_o$ is given by
\begin{align}
t^o = \frac{P}{\sigma^2 \lambdabf_{\text{\tiny QoS}}^{oT}\gammabf}.
\end{align}
\end{theorem}

\IEEEproof
Using the equivalent problem $\Sc_1(\gammabf,P)$, and from  \eqref{eqn:inv_PtoS} and \eqref{eqn:inv_StoP}, we first consider the inverse QoS problem $\Pc_o(t^o\gammabf)$. By Corollary~\ref{cor2}, the minimum power of  $\Pc_o(t^o\gammabf)$ is
\begin{align*}
P = t^o\sigma^2 \lambdabf_{\QoS}^{oT}\gammabf
\end{align*}
where $\lambdabf_{\QoS}^o$ is  obtained in \eqref{w_o2}  for  $\Pc_o(t^o\gammabf)$. Thus, $t^o = P/(\sigma^2\lambdabf_{\QoS}^{oT}\gammabf)$. Based on the inversion relation in \eqref{eqn:inv_PtoS},  the optimal  $\wbf_{\MMF,i}^o$ for  $\Sc_1(\gammabf,P)$ is the same as in \eqref{w_o2}, except that  $\gamma_{ik}$ in \eqref{w_o2} is now replaced by $t^o\gamma_{ik}=\gamma_{ik}P/(\sigma^2\lambdabf_{\QoS}^{oT}\gammabf)$. Correspondingly, using $t^o\gamma_{ik}$ and $\lambdabf_{\QoS}^o$, the covariance matrix $\Rbf(\lambdabf^o)$ in  \eqref{w_o2} becomes $ \widetilde{\Rbf}(\lambdabf_\text{\tiny QoS}^o)$  in \eqref{R_tilde}, and $a_{ik}^o$ in  \eqref{w_o2} now becomes $\tilde{a}_{ik}^o$ shown in \eqref{a_MMF}.
Thus, we have the optimal $\wbf_{\MMF,i}^o$ given in \eqref{w_o2_MMF}.
\endIEEEproof

Similar to Theorem~\ref{thm1} for the QoS problem, the optimal  beamforming vector $\wbf_{\MMF,i}^o$ for the MMF problem in Theorem~\ref{thm2} is in  semi-closed-form as a function of $t^o,\lambdabf_{\QoS}^o$, and $\tilde{\abf}_i^o$. The expression in \eqref{w_o2_MMF} provides the optimal MMF beamforming  structure. We  still need to numerically determine  $t$,   $\lambdabf_{\QoS}$ related to the QoS problem $\Pc_o(t\gammabf)$, and   $\tilde{\abf}_i$, which are difficult to compute. As mentioned earlier, using the inverse problem relationship, one practical method to  obtain $\{\wbf_{\MMF,i}^o\}$ is through iteratively finding $\{\wbf^o_i\}$ for $\Pc_o(t\gammabf)$ with a bi-section search over $t$ until the transmit power is equal to $P$.
Since this procedure is known in the literature, details are omitted.

\subsection{Asymptotic MMF Multicast Beamforming Solution}
The difficulty of directly computing  $\wbf_{\MMF,i}^o$ in \eqref{w_o2_MMF}  is in the determination of $\widetilde{\Rbf}(\lambdabf_\text{\tiny QoS}^o)$, because it requires the knowledge of $t^o$. Note from \eqref{R_tilde} that the contribution from each user channel is weighted by $\frac{P}{\sigma^2} \cdot \frac{\lambda_{\text{\tiny QoS},ik}^o\gamma_{ik}}{\lambdabf_\text{\tiny QoS}^{oT}\gammabf}$, indicating the fraction of  transmit power used by each user.
For massive MIMO systems with large $N$, we may obtain an asymptotic expression for $\widetilde{\Rbf}(\lambdabf_\text{\tiny QoS}^o)$   and consider a simplified fast computation method. Specifically, we  use  the asymptotic expression of $\lambdabf_{\QoS}^o$ in Proposition~\ref{prop3} to obtain the asymptotic expression for $\widetilde{\Rbf}(\lambdabf_\text{\tiny QoS}^o)$. Consider each channel as $\hbf_{ik}=\sqrt{\beta_{ik}}\gbf_{ik}$. As an example, in the special case $\gamma_{ik}=\gamma$, $\forall i,k$,  using the first term in \eqref{prop3:gamma}  to approximate $\lambda^o_{\QoS,ik}\beta_{ik}$, we can approximate $\widetilde{\Rbf}(\lambdabf_\text{\tiny QoS}^o)$  for large $N$ using its simple asymptotic expression given by
\begin{align} \label{R_tilde:simple}
\widetilde{\Rbf}(\lambdabf_\text{\tiny QoS}) \approx \Ibf + \frac{P}{\sigma^2K_\tot}\bar{\beta}_\text{h} \sum_{i=1}^G \sum_{k=1}^{K_i}  \gbf_{ik}\gbf_{ik}^H \triangleq \widetilde{\Rbf}^{\scriptscriptstyle \infty}_\MMF
\end{align}
where
$\bar{\beta}_\text{h}\triangleq  1/(\frac{1}{K_\tot}\sum_{i=1}^G\sum_{k=1}^{K_i} \frac{1}{\beta_{ik}})$ is the harmonic mean of the large-scale channel variations of all users. As the asymptotic $\widetilde{\Rbf}^{\scriptscriptstyle \infty}_\MMF$ in \eqref{R_tilde:simple} is in closed-form, we only need to compute weight vector $\tilde{\abf}_i$ in \eqref{w_o2_MMF} to obtain $\wbf_{\MMF,i}$. Similar to  Section~\ref{subsec:weight_a}, using  \eqref{w_o2_MMF}, we can transform  $\Sc_1$ into the weight optimization problem w.r.t. $(\{\tilde{\abf}_i\},t)$ of a much smaller size.  The SDR or SCA method can be similarly applied, along with a bi-section search over $t$, to obtain a solution.

To further simplify the computation of $\wbf^o_{\MMF,i}$, we also propose a closed-form asymptotic beamformer $\wbf^{\scriptscriptstyle \infty}_{\MMF,i}$, where besides using $\widetilde{\Rbf}^{\scriptscriptstyle \infty}_\MMF$ in \eqref{R_tilde:simple}, we replace  weight vector  $\tilde{\abf}_i$ by its asymptotic version   $\tilde{\abf}^{\scriptscriptstyle \infty}_i$. The asymptotic weight $\tilde{\abf}^{\scriptscriptstyle \infty}_i$ has been obtained in the limiting regime  $N\to \infty$, when all interferences vanish \cite{Zhou&Tao:ICC15} (effectively each group becomes a separate single-group scenario). The expression of   $\tilde{\abf}^{\scriptscriptstyle \infty}_i$ is given by
$\tilde{\abf}^{\scriptscriptstyle \infty}_i =c_i \qbf_i$, where $\qbf_i=[1/\beta_{i1},\ldots,1/\beta_{iK_i}]^T$, and $c_i$ is the scaling factor for $\tilde{\abf}^{\scriptscriptstyle \infty}_i$ to ensure that the transmit power allocated to group $i$ is  $\|\wbf^{\scriptscriptstyle \infty}_{\MMF,i}\|^2=\frac{K_i\bar{\beta}_\text{h}}{K_\tot\bar{\beta}_{\text{h},i}}P$,
with $\bar{\beta}_{\text{h},i}\triangleq  1/(\frac{1}{K_i}\sum_{k=1}^{K_i} \frac{1}{\beta_{ik}})$ being the harmonic mean of $\{\beta_{ik}\}$ for users in group $i$. Using the above, we have the  proposed asymptotic MMF multicast beamformer
in the following simple closed-form expression
\begin{align}\label{w_MMF_asym}
\wbf^{\scriptscriptstyle \infty}_{\MMF,i} = \widetilde{\Rbf}^{{\scriptscriptstyle \infty}-1}_\MMF\Hbf_i\tilde{\abf}^{\scriptscriptstyle \infty}_i, \ i\in \Gc.
\end{align}
By \eqref{w_MMF_asym}, we obtain the scaling factor $c_i$ for $\tilde{\abf}^{\scriptscriptstyle \infty}_i$, given by $c_i^2 =  \frac{\bar{\beta}_{\text{h},i}}{\bar{\beta}_\text{h}}P/(\qbf_i^H\Hbf_i^H\widetilde{\Rbf}^{{\scriptscriptstyle \infty}-2}_\MMF\Hbf_i\qbf_i)$.

\vspace*{0.5em}
\noindent{\bf Remark}: We point out that our asymptotic beamforming solution in \eqref{w_MMF_asym} is different from other existing asymptotic beamformers  in the literature \cite{Zhou&Tao:ICC15,Sadeghi&Yuen:Globecom15}. They are identical when $N\to \infty$. However, for finite large $N$, $\wbf^{\scriptscriptstyle \infty}_{\MMF,i}$ contains the interference term in $\widetilde{\Rbf}^{\scriptscriptstyle \infty}_\MMF$, while existing asymptotic beamformers ignore interference. For this reason, as verified in simulation, our asymptotic beamformer converges to the optimal beamformer much faster, around $N \approx 500$. In contrast, existing ones require $N$ to be more than a few thousand.

\section{Simulation Results} \label{sec:sim}

We consider a symmetric setup for downlink multi-group multicast beamforming, where  $K_i=K$, $\forall i$, and the target received SINR  $\gamma_{ik} =\gamma$, $\forall k,i$.
Unless otherwise specified, we
set the default system setup as  $G=3$ groups, $K=5$ users per group, and  $\gamma=\ 10$ dB.  Channel vectors  are generated  i.i.d. as $\hbf_{ik}\sim\Cc\Nc( {\bf 0},\beta_{ik}\Ibf)$, $\forall k,i$. We consider two types of channels: 1) pathloss channels:  $\beta_{ik}=\xi_od_{ik}^{-3}$, where  $d_{ik}$ is the  distance between the BS and  user $k$ in group $i$, generated randomly, pathloss exponent is  3, and $\xi_o$ is the pathloss constant. We set $\xi_o$ such that at the cell boundary, the nominal average received SNR (by a single transmit antenna and unit transmit power)  is $-5$dB; 2) normalized channels: for all users, $\beta_{ik}=1$, $\forall k,i$, \ie  all users having the same distance to the BS. The performance results  are obtained by averaging over 100 channel realizations per user (also  over 10 realizations of user locations for pathloss channels).

\subsubsection{ Convergence behavior of the algorithm for $\lambdabf$}
\begin{figure}[t]
        \centering
        \includegraphics[width=.55\columnwidth]{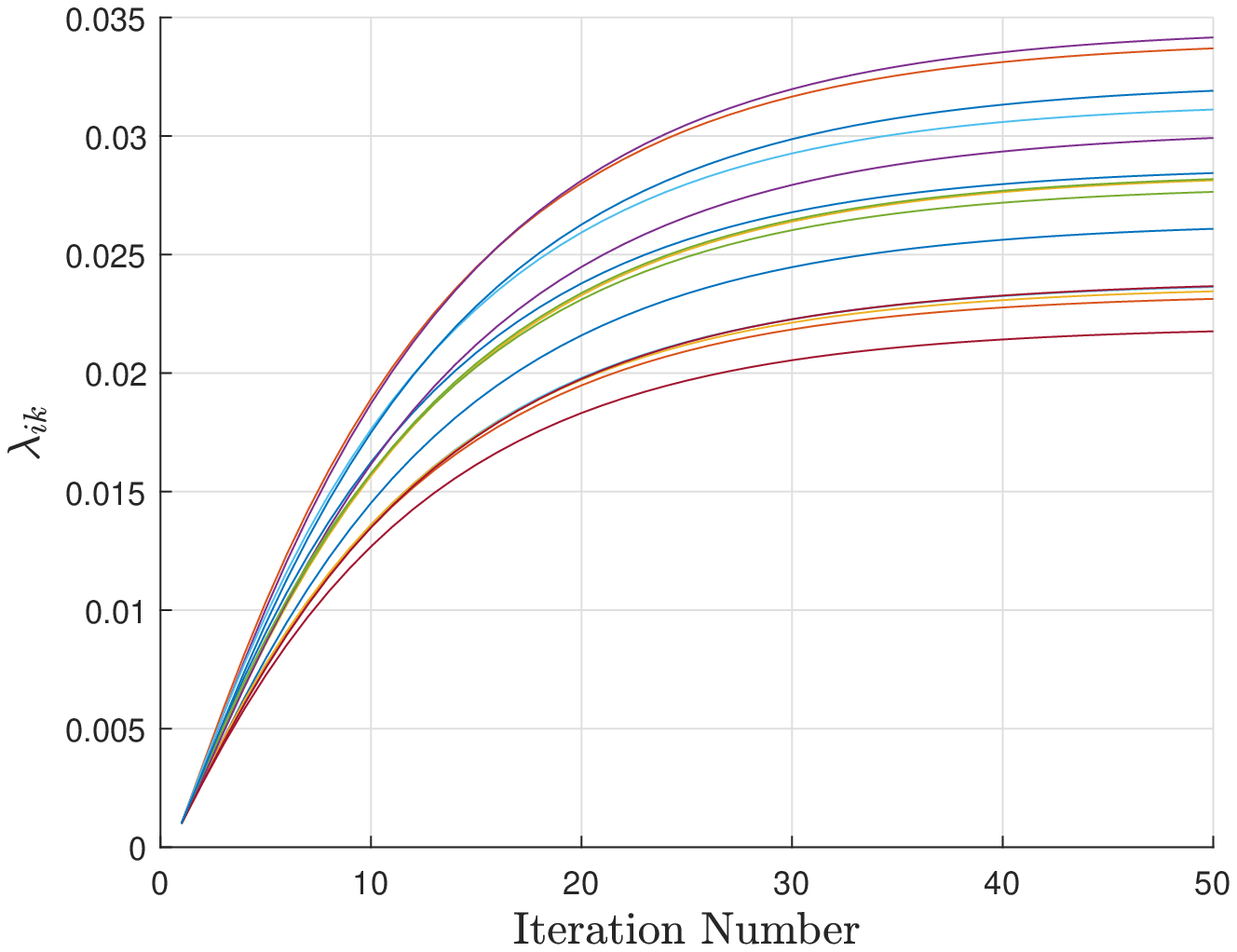}\hspace*{-1.1em}\includegraphics[width=.55\columnwidth]{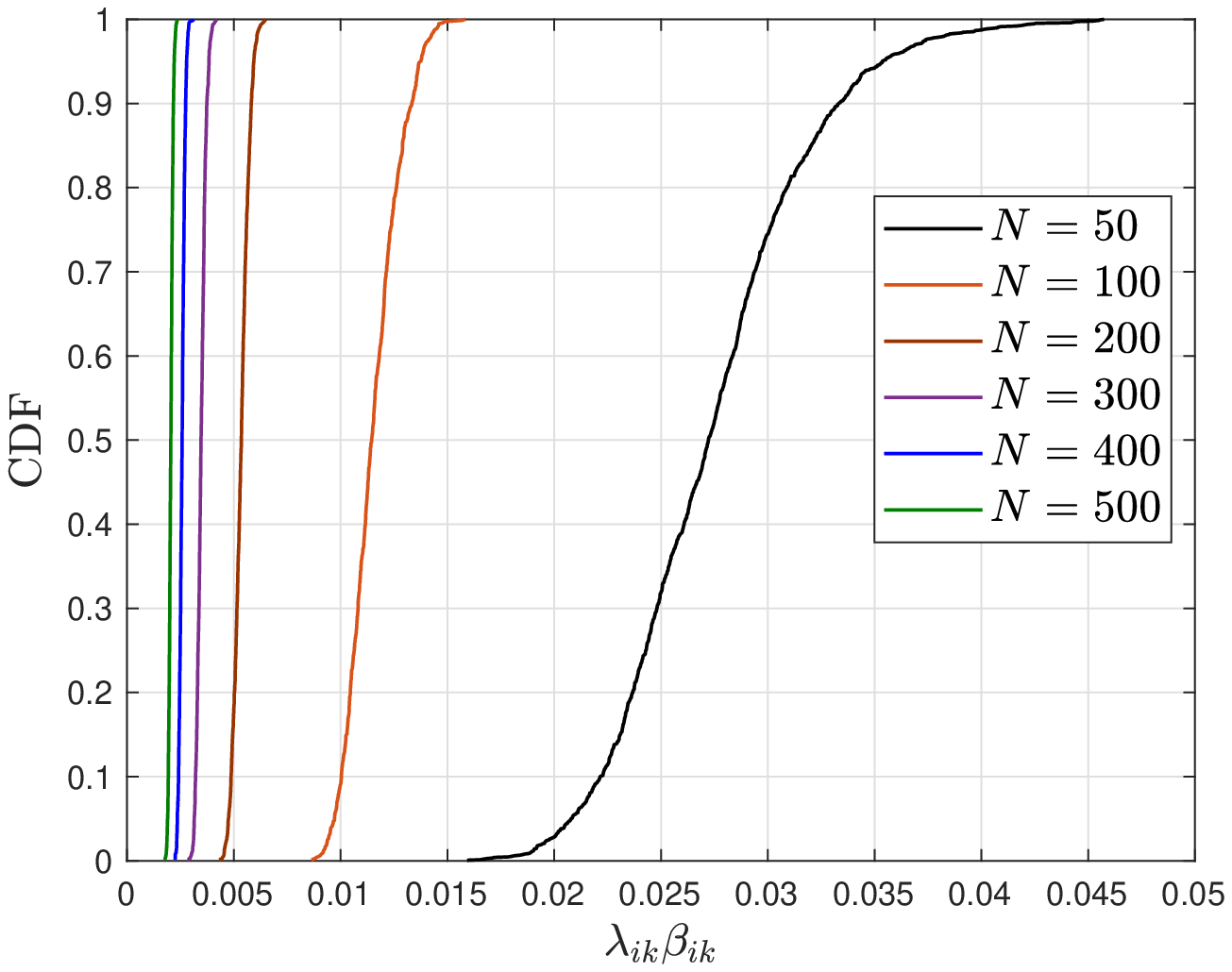}        \centering
        \caption{Left: Convergence behavior of the iterative algorithm for $\{\lambda_{ik}\}$   ($N=50$, $G=3$, $K=5$). Each curve represents $\lambda_{ik}$ of a user  over iterations. Right: The CDF of $\lambda_{ik}\beta_{ik}$  for randomly generated user locations.} \label{fig:lambda_alg} \vspace*{-1em}
\end{figure}

We first study the convergence behavior of the  iterative algorithm proposed in Section~\ref{subsec:lambda_alg} to compute $\lambdabf$  for the QoS problem  $\Pc_o$. Fig.~\ref{fig:lambda_alg} (left) shows the trajectory of  $\lambda_{ik}$  over the number of iterations for each user with a normalized channel, for  $N=50$.
To verify the asymptotic behavior of $\lambda_{ik}\beta_{ik}$ in Proposition~\ref{prop3},  we consider users randomly located in the cell and generate their pathloss channels accordingly. Fig.~\ref{fig:lambda_alg} (right) shows the CDF of $\lambda_{ik}\beta_{ik}$, with  $\lambda_{ik}$ being computed by the iterative algorithm, for $N=50$ to 500. It is evident that as $N$ becomes large,  all $\lambda_{ik}\beta_{ik}$'s converge to the same value, and the CDF converges to a step function.

\subsubsection{Performance comparison for the QoS problem}
We present the performance of using the optimal beamforming structure $\wbf_i^o$ in \eqref{w_o2} and our proposed algorithms for the QoS problem $\Pc_o$. Both  SDR and SCA methods in Section~\ref{subsec:weight_a}  are considered for computing  weight vector $\abf_i$, and we name them OptBFwSDR and OptBFwSCA, respectively. Normalized channels are used. We also consider the following  for comparison: 1) Lower bound for  $\Pc_o$: obtained by solving the relaxed problem of $\Pc_o$ via SDR, it serves as a benchmark for all algorithms; 2) AsymBFwSCA: the same as OptBFwSCA, except that  $\Rbf(\lambdabf)$ is approximated by \eqref{prop3:R};  3)  Direct SDR:  directly solve $\Pc_o$ for $\wbf$   via  SDR   with Gaussian randomization; 4) Direct SCA: directly solve $\Pc_o$ for $\wbf$   via the SCA  method,  taking the solution from the direct SDR  as the initial point; 5) BDZF\cite{Sadeghi&etal:TWC17}: a  low-complexity  algorithm  proposed recently for large-scale antenna arrays, using a two-layered approach combining block-diagonal ZF beamforming and SCA.\footnote{Due to ZF beamforming, BDZF requires $N> K_\tot-\min_{i\in\Gc} K_i$.} The computational complexity of SDR or SCA-based algorithm is analyzed in Sections~\ref{subsec:weight_a_SDR} and \ref{subsec:weight_a_SCA}, respectively. The complexity of BDZF is $\Oc(GN^3)$ in each SCA iteration for $N\gg K$.
\begin{figure}[t]
\centering
\includegraphics[width=.65\columnwidth]{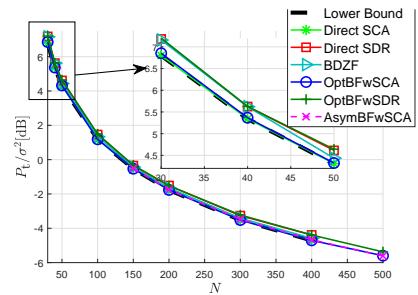}\\[-.5em]
\caption{ QoS: Normalized transmit power $P_t/\sigma^2$  vs. $N$  ($G=3$, $K=5$).}\label{fig:PowervsM}
\vspace{-.5em}
\end{figure}
\begin{table}[t]
\renewcommand{\arraystretch}{1}
\centering
\caption{ Average Computation Time over $N$ (sec.) ($G=3,K=5$, QoS).}
\label{tab:vsM}
\vspace*{-.5em}
\begin{tabular}{p{1.5cm}||p{0.5cm}|p{0.5cm}|p{0.5cm}|p{0.5cm}|p{0.5cm}|p{.5cm}}
\hline
$N$ & 50 & 100 & 200 & 300 & 400 & 500\\
\hline\hline
OptBFwSDR & 0.49 & 0.45 & 0.49 & 0.52 & 0.56 & 0.62\\
\hline
OptBFwSCA &1.61 & 1.37 & 1.51 & 1.40 & 1.41 & 1.42\\
\hline
BDZF\cite{Sadeghi&etal:TWC17} &11.5 & 34.1 & 182 & 495 & 605 & N/A\\
\hline
Direct SDR  & 8.7 & 52.9 & 427 & 1509 & 4507 & N/A\\
\hline
Direct SCA & 7.41 & 44.2& 353 & 1192 & N/A & N/A \\ \hline
\end{tabular}
\end{table}
\begin{figure}[!h]
\centering
\includegraphics[width=0.65\columnwidth]{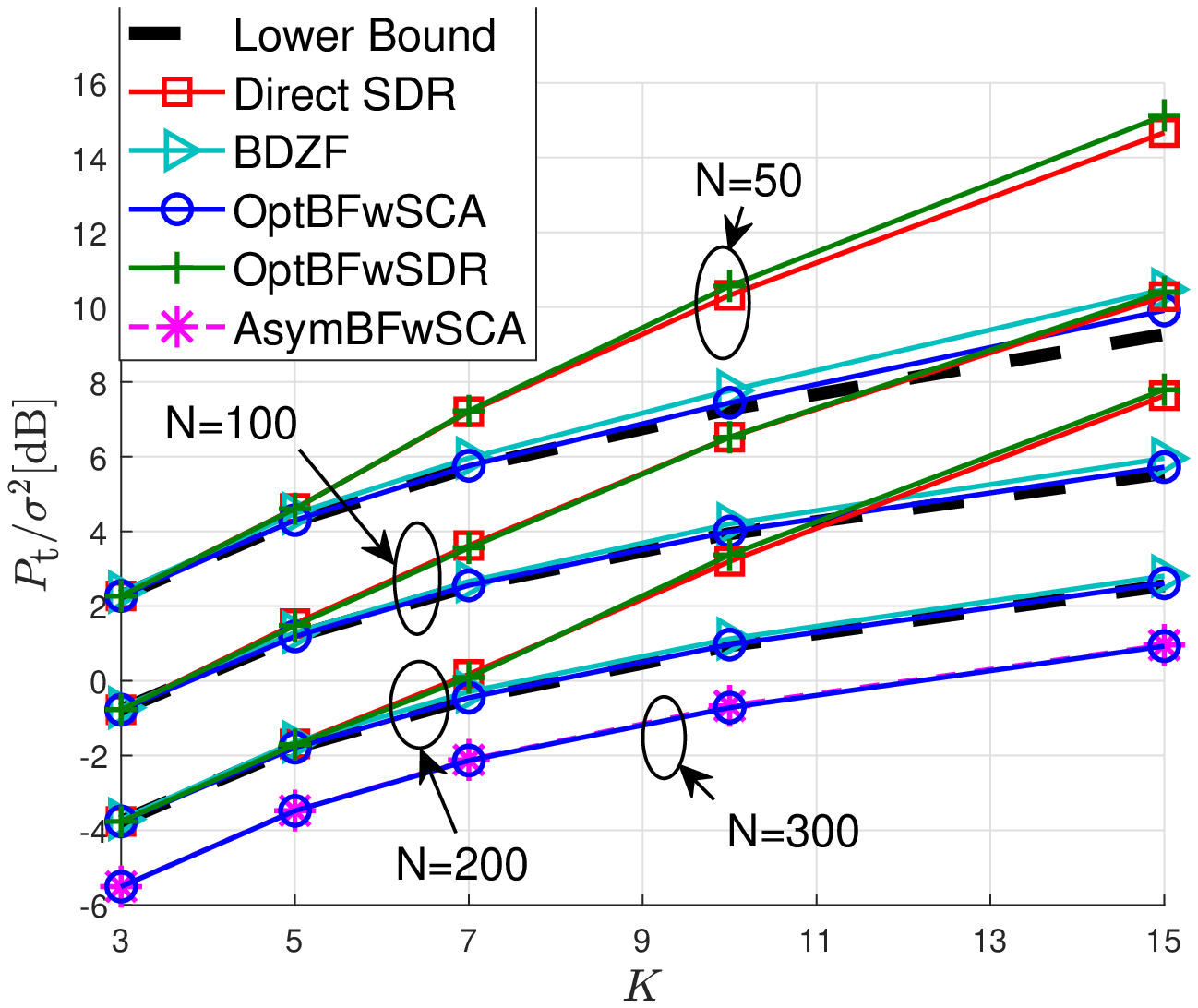}
\caption{QoS: Normalized transmit power $P_t/\sigma^2$  vs. $K$  ($G=3$).} \label{fig:PowervsK} 
\vspace{-.5em}
\end{figure}
\begin{table}[!h]
\centering
\caption{ Average Computation Time over $K$  (sec.) ($N=100, G=3$, QoS).}
\label{tab:vsK}
\vspace*{-.5em}
\begin{tabular} {p{1.5cm}||p{.5cm}|p{.5cm}|p{.5cm}|p{.5cm}|p{.5cm}}
\hline
$K$& 3 & 5 & 7 & 10 & 15 \\
\hline\hline
OptBFwSDR & 0.44 & 0.48 & 0.68 & 1.03 & 1.89\\
\hline
OptBFwSCA & 0.81 & 1.50 & 3.28 & 6.46 & 13.07\\
\hline
BDZF\cite{Sadeghi&etal:TWC17} & 21.9     & 29.2 & 38.0 & 46.9 & 51.2\\
\hline
Direct SDR & 33.5 & 50.6 & 69.4 & 98.4 & 136\\
\hline
\end{tabular}
\end{table}
\begin{figure}[t]
\centering
\includegraphics[width=0.65\columnwidth]{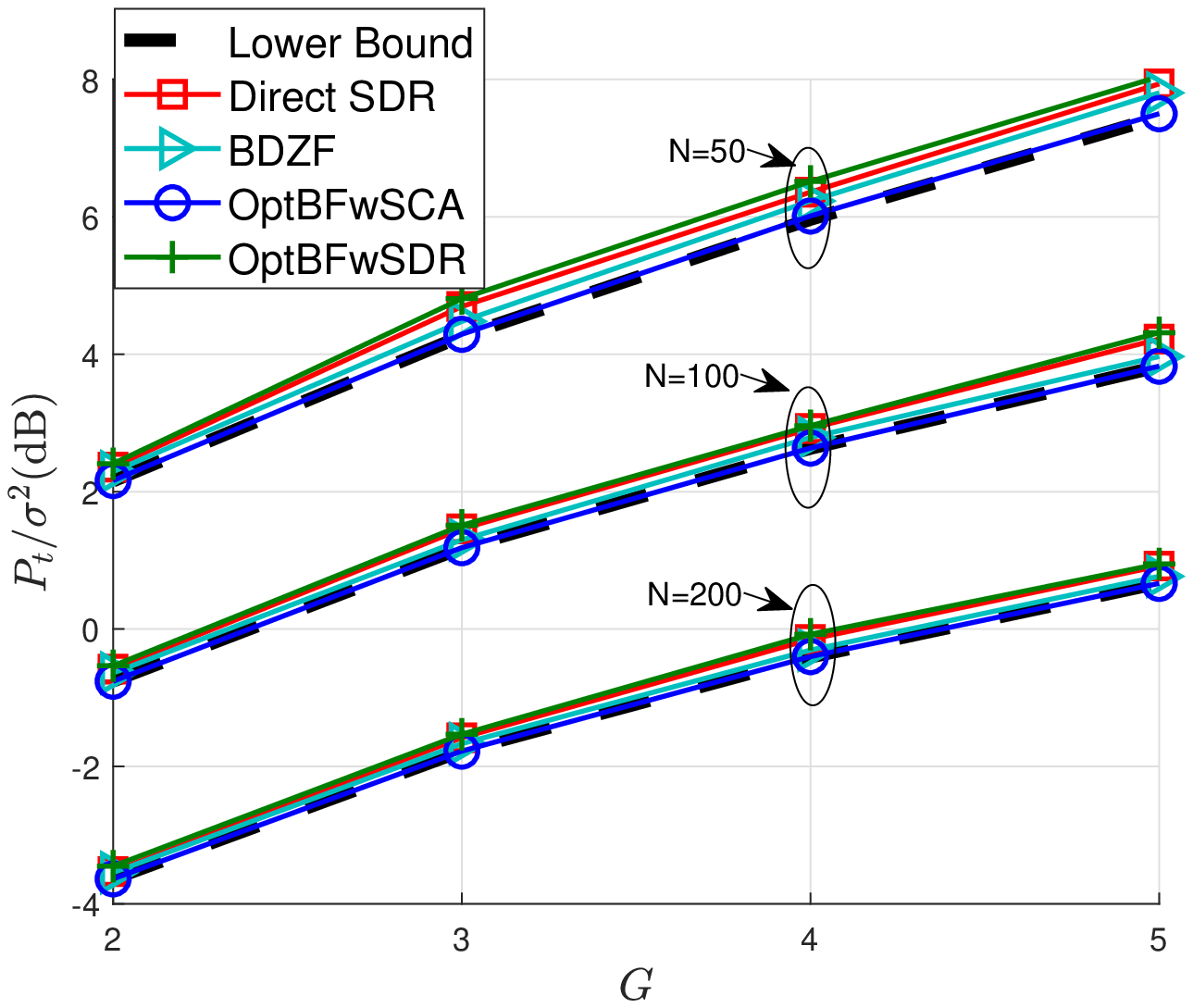}\\[-.5em]
\caption{QoS: Normalized transmit power $P_t/\sigma^2$  vs. $G$  ($K=5$).} \label{fig:PowervsG} 
\end{figure}
\begin{table}
\centering
\caption{Average Computation Time over $G$  (sec.) ($N=100, K=5$).}
\label{tab:vsG}
\begin{tabular} {p{2.2cm}||p{.8cm}|p{.8cm}|p{.8cm}|p{.8cm}}
\hline
$G$& 2 & 3 & 4 & 5 \\
\hline\hline
OptBFwSDR & 0.39 & 0.48 & 0.63 & 0.80\\
\hline
OptBFwSCA & 0.86 & 1.50 & 2.77 & 4.52\\
\hline
BDZF\cite{Sadeghi&etal:TWC17} & 16.5 & 27.0 & 40.2 & 49.2\\
\hline
 Direct SDR & 22.6 & 50.6 & 86.4 & 132.6\\
\hline
\end{tabular}
\vspace*{-1em}
\end{table}

  Denote the transmit power objective  of $\Pc_o$ by  $P_t\triangleq \sum_{i=1}^G\|\wbf_i\|^2$.
Fig.~\ref{fig:PowervsM} shows the average normalized  transmit power $P_t/\sigma^2$ vs. the number of antennas $N$.  Both OptBFwSCA and OptBFwSDR have consistent performance  over a wide range of  $N$ values. The performance of   OptBFwSCA   nearly attains the lower bound, while that of   OptBFwSDR has a small gap of $\sim 0.3$ dB. Their performance is near-identical to their respective direct methods (direct SCA or direct SDR). The computational saving  by using the optimal beamforming structure in OptBFwSCA and OptBFwSDR is  evident in Table~\ref{tab:vsM}, where the average computation time    for different $N$ is shown (via MATLAB and CVX). Both   OptBFwSCA and OptBFwSDR  require very low computation time, which is   kept roughly constant for all $N$ values. This is in contrast to the other alternative methods,  whose computation times  increase fast with $N$ and  become  impractical for large $N$. OptBFwSCA performs better than OptBFwSDR by using SCA, at the cost of slightly higher computational complexity. Furthermore, we observe that  AsymBFwSCA performs nearly identical to OptBFwSCA, indicating the effectiveness by using the closed-form asymptotic expression in \eqref{prop3:R} for $\Rbf(\lambdabf)$.

Fig.~\ref{fig:PowervsK} presents the average normalized  transmit power $P_t/\sigma^2$ vs.  $K$ users per group  for different $N$ values, when $G=3$. OptBFwSCA performs very well and nearly attains the lower bound at all $K$ and $N$ values. For both OptBFwSDR  and the direct SDR, the performance deteriorates over $K$, which is known  for the SDR-based methods as the number of constraints ($K$)  becomes large. The average computation time  for the plots in  Fig.~\ref{fig:PowervsK} is shown in Table~\ref{tab:vsK}.
The increase of computation time  over $K$ by  OptBFwSDR  is insignificant, while that of  OptBFwSCA  is more noticeable. Nevertheless, the computation time under both methods is still kept very low and is significantly lower than other methods. Also, to verify the performance of AsymBFwSCA, we plot it over $K$ for $N=300$. Again, it shows a near-identical performance as OptBFwSCA. Fig.~\ref{fig:PowervsG} shows the average normalized  transmit power  $P_t/\sigma^2$  vs.  $G$ groups when $K=5$, for different $N$ values. The corresponding average computation time is shown in Table~\ref{tab:vsG}. The relative performance among different methods  is similar to that  in Fig.~\ref{fig:PowervsK} for $K=5$ and $G=3$ and maintains the same as $G$ increases: both OptBFwSDR and OptBFwSCA provide near-optimal performance with very low computation time that only slightly increases over $G$. 

\begin{figure}[t]
        \centering
        \includegraphics[width=.65\columnwidth]{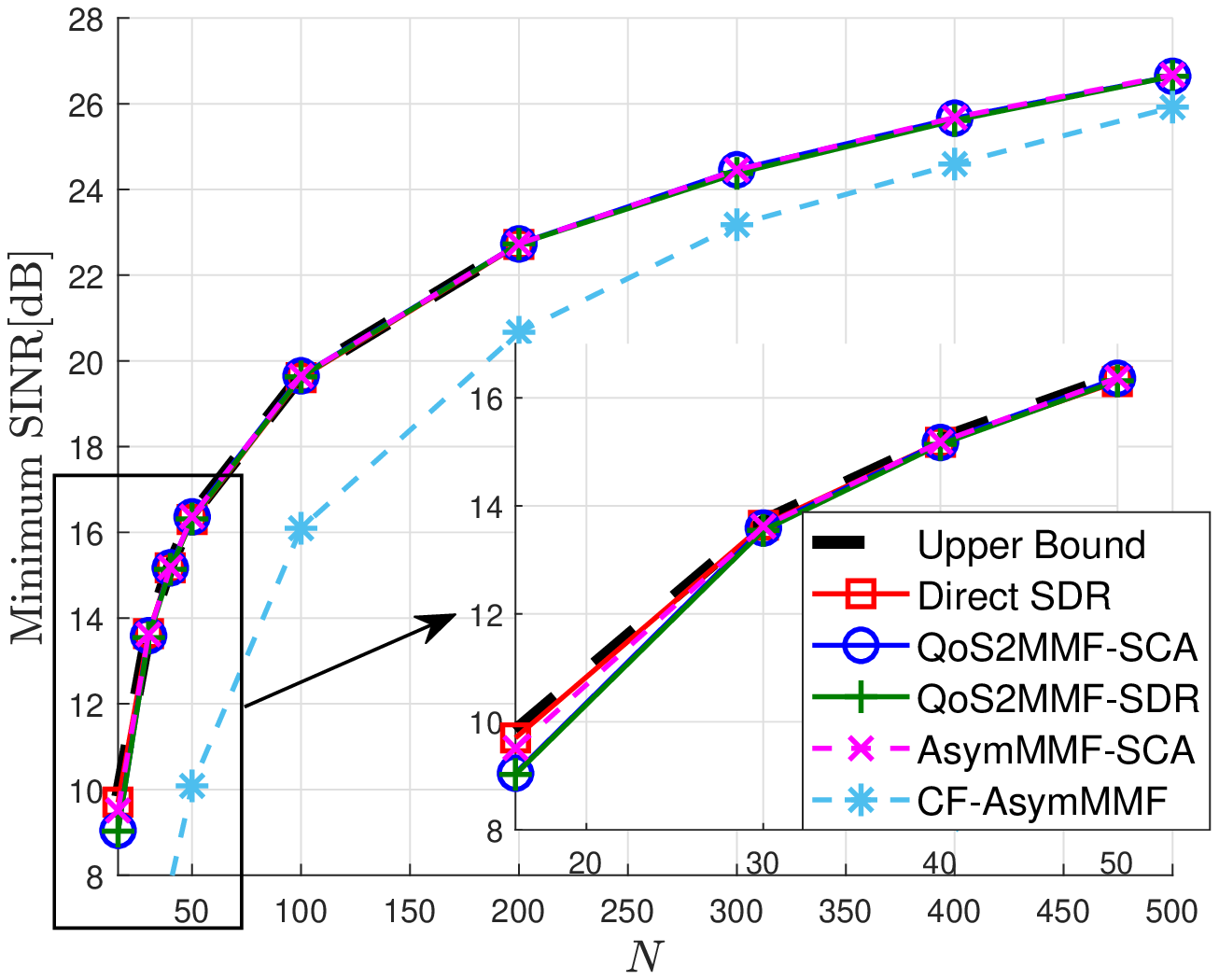}\\[-.5em]
        \centering
        \caption{MMF: Minimum SINR vs. $N$ ($G=3$, $K=5$).}
        \label{fig:SINRvsN}
        \vspace{.5em}
        \centering
        \includegraphics[width=0.65\columnwidth]{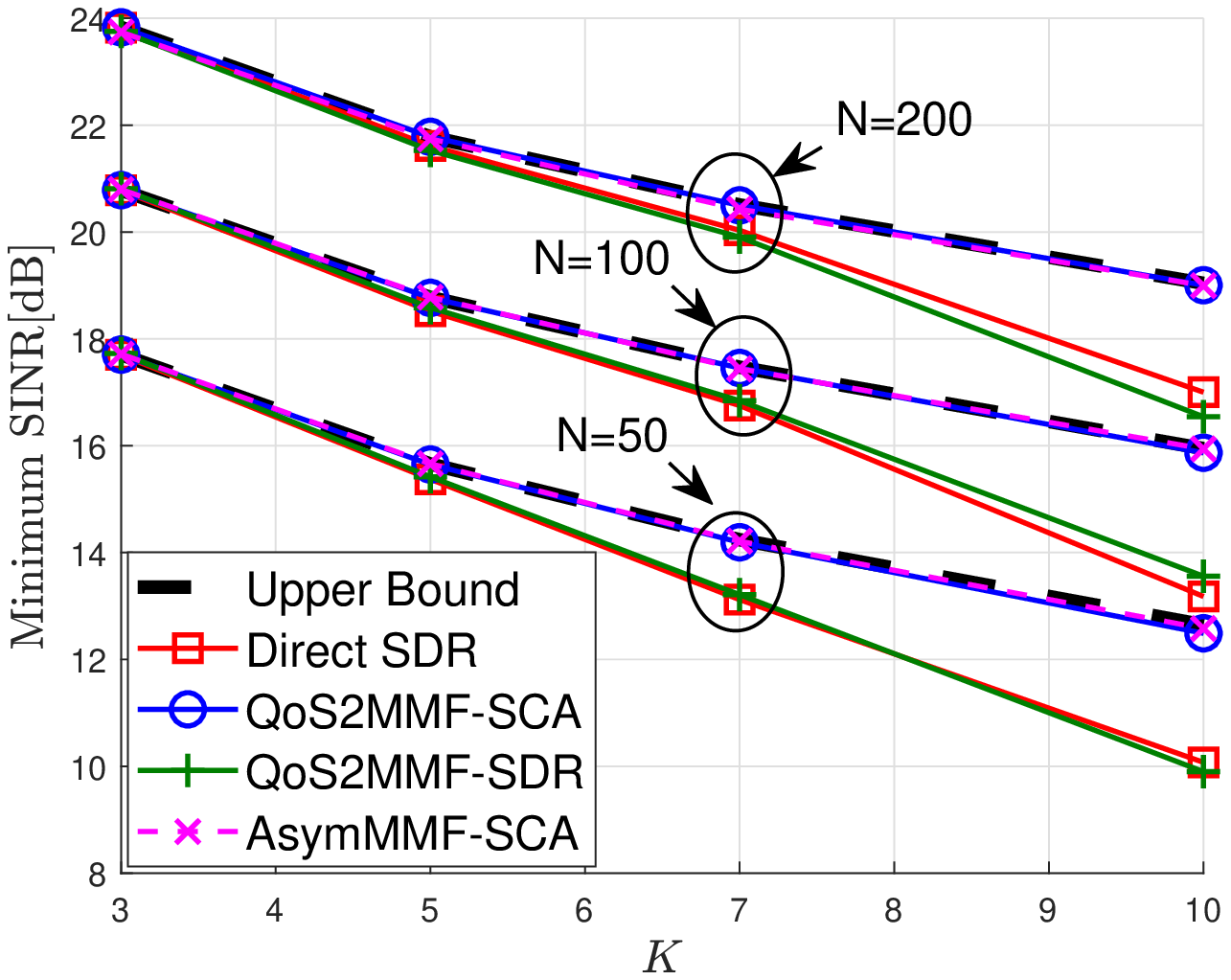}\\[-.5em]
        \centering
        \caption{MMF: Minimum SINR  vs. $K$ ($G=3$).}
        \label{fig:SINRvsK} \vspace*{-0.5em}
\end{figure}

\subsubsection{Performance comparison for the MMF problem} We now present the performance using the optimal solution  structure in \eqref{w_o2_MMF} for the MMF problem  $\Sc_o$.
The MMF beamformer is obtained via iteratively solving the QoS problem discussed in Section~\ref{sec:MMF}, and the weight $\tilde{\abf}_i$ is computed via the SDR (SCA) method,  which we refer to as QoS2MMF-SDR (QoS2MMF-SCA).  The pathloss channels are used. For comparison, we also consider the followings: 1) Upper bound of $\Sc_o$: obtained by solving the relaxed version  of $\Sc_1$via SDR; 2) Direct SDR: direct solve the relaxed version of $\Sc_1$ via SDR with Gaussian randomization;\footnote{In both 1) and 2), a bi-section search over $t$ is required along with SDR to obtain a solution.} 3) AsymMMF-SCA: approximate  $\widetilde{\Rbf}(\lambdabf_\text{\tiny QoS})$  by \eqref{R_tilde:simple} and compute weight vector $\tilde{\abf}_i$ by the SCA method as described below \eqref{R_tilde:simple}; 4) CF-AsymMMF:  the closed-form asymptotic beamformer given in \eqref{w_MMF_asym}.

\begin{table}[!h]
\renewcommand{\arraystretch}{1}
\centering
\caption{ Average Computation Time over $N$ (sec.) ($G=3,K=5$, MMF).}
\label{tab:QoS2MMFvsM}
\vspace*{-.5em}
\begin{tabular} {p{2cm}||p{0.6cm}|p{0.6cm}|p{0.6cm}|p{0.6cm}|p{0.6cm}|p{0.6cm}}
 \hline
 $N$& 50 & 100 & 200& 300&400&500\\
 \hline\hline
  QoS2MMF-SDR &6.77&7.26&7.54&7.75&8.88&9.96\\
 \hline
 QoS2MMF-SCA &27.6&29.9&27.6&30.0&33.7&37.6\\
 \hline
 AsymMMF-SCA&17.4 &17.8  &18.4   &17.2      &17.8&17.1\\
 \hline
  Direct SDR &162&1158&9924&N/A&N/A&N/A\\
 \hline
 \end{tabular}
 \vspace*{1.5em}
 \renewcommand{\arraystretch}{1}
 \centering
 \caption{ Average Computation Time over $K$  (sec.) ($N=100, G=3$, MMF).}
        \label{tab:QoS2MMFvsK}
        \vspace*{-.5em}
\begin{tabular} {p{2cm}||p{.6cm}|p{.6cm}|p{.6cm}|p{.6cm}}
 \hline
 $K$& 3 & 5 & 7 & 10 \\
\hline\hline
  QoS2MMF-SDR &5.65&7.16&8.81&14.6\\
  \hline
   QoS2MMF-SCA &18.3&22.7&27.7&39.3\\
    \hline
 AsymMMF-SCA& 7.21 & 10.7 & 16.2 & 30.3\\
    \hline
  Direct SDR &457&766&1018&1359\\
  \hline
\end{tabular}
\vspace*{-1em}
\end{table}

Fig.~\ref{fig:SINRvsN} shows the average minimum SINR vs. $N$, and Table~\ref{tab:QoS2MMFvsM} shows the corresponding computation time. The observations are similar to that in the QoS problem, where both QoS2MMF-SCA and QoS2MMF-SDR provide  near-optimal performance, with a substantially lower computation time that only increases slightly over $N$. Furthermore, AsymMMF-SCA performs as good as QoS2MM-SCA,   with a further lower computation time   roughly constant over $N$. This verifies the asymptotic expression of $\widetilde{\Rbf}(\lambdabf_\text{\tiny QoS})$ in  \eqref{R_tilde:simple} and  the effectiveness of this efficient method for the MMF problem. In contrast,  CF-AsymMMF converges to the upper bound slower over $N$, with a more noticeable performance gap observed due to the simple closed-form asymptotic weights used. Nonetheless, it offers much better performance  with a significantly improved  convergence rate than the existing asymptotic beamformers \cite{Zhou&Tao:ICC15,Sadeghi&Yuen:Globecom15},  with less than 1dB gap at $N=500$ (as compared to $N$ being a few thousands in those works).

Finally, Fig.~\ref{fig:SINRvsK} shows the average minimum SINR vs. $K$, with the computation time shown in  Table~\ref{tab:QoS2MMFvsK}. The observations are similar to that in the QoS problem, where both  QoS2MMF-SCA and AsymMMF-SCA show near-optimal performance at different $K$ values, and the computation time of the proposed methods is substantially lower than the direct SDR method.

\section{Conclusion and Discussion}\label{sec:con}
In this work, we obtained the optimal beamforming  structure for the multi-group multicast beamforming, which has been unknown in the literature. Combining both the SCA numerical method and  Lagrange duality, we derived the optimal multicast beamforming  structure for both the QoS and MMF problems. This structure sheds light on the optimal multicast beamforming: 1) There is an uplink-downlink duality interpretation for the multicast beamforming problem, similar to the classical downlink multi-user unicast beamforming problem. 2) The optimal multicast beamformer is a weighted MMSE filter based on a group-channel direction, as a generalized version of the optimal downlink unicast beamformer. 3) There is an inherent low-dimensional structure in the optimal beamforming solution independent of $N$, which brings opportunities for efficient numerical algorithms to compute the beamformer that is especially beneficial for systems with large antenna arrays. Using the optimal beamforming structure, we proposed efficient algorithms to compute the parameters in the optimal multicast beamformer. Characterizing the asymptotic behavior of the beamformers as $N$ grows large, we provided simple   approximate multicast beamformers for large $N$, including a closed-form asymptotic beamformer. They provide  practical multicast beamforming solutions with near-optimal performance at very low  computational complexity   for massive MIMO systems.

The optimal multicast beamforming structure can be extended to multi-cell coordinated multicast beamforming scenarios, where the model difference is  the per BS transmit power, instead of total power. For example, the MMF problem $\Sc_o$ has an inherent power allocation problem to $G$ groups, while for the multi-cell problem, the transmit power at each BS (to each group) is fixed. However, this difference is not expected to fundamentally change the optimal multicast beamforming  structure, and the results obtained in this work can be extended to the multi-cell problem after some care of technical details.
\appendices
\section{Proof of Corollary~\ref{cor2}}\label{app:cor2}
\IEEEproof
Since  $\Pc_\SCA(\zbf)$ is convex, its minimum objective can be obtained by its dual $\Dc_\SCA(\zbf)$. Rewrite $\nubf_i$ defined above \eqref{Sigma} in a compact matrix form as $\nubf_i = \Hbf_i\Dbf_{\lambdabf_i}\Hbf_i^H\zbf_i$, where $\Dbf_{\lambdabf_i}\triangleq\diag(\lambdabf_i)$.
Then, the optimal $\wbf_i^\star(\zbf)$ in \eqref{w_o_SCA} can be rewritten as\\[-1.5em]
\begin{align}
\wbf_i^\star(\zbf) &=\Rbf_{i^-}^{-1}(\lambdabf)\Hbf_i\Dbf_{\lambdabf_i}\Hbf_i^H\zbf_i.
\end{align}
Substituting the above expression into \eqref{multi:newdual}, we have the dual function $g(\zbf,\lambdabf)$ in \eqref{dual_func} as\\[-1.5em]
\begin{align}
g(\zbf,\lambdabf) =& \sum_{i=1}^G\sum_{k=1}^{K_i} \lambda_{ik}(\sigma^2\gamma_{ik}+|\zbf_i^H\hbf_{ik}|^2) \nn\\
&+ \sum_{i=1}^G\zbf_i^H\Hbf_i\Dbf_{\lambdabf_i}\Hbf_i^H\Rbf_{i^-}^{-1}(\lambdabf)\Hbf_i\Dbf_{\lambdabf_i}\Hbf_i^H\zbf_i \nn\\
&-2\sum_{i=1}^G\Re\{\zbf_i^H\Hbf_i^H\Dbf_{\lambdabf_i}\Hbf_i\Rbf_{i^-}^{-1}(\lambdabf)\Hbf_i\Dbf_{\lambdabf_i}\Hbf_i^H\zbf_i\}\nn\\
=&\sigma^2 \sum_{i=1}^G\lambdabf_i^T\gammabf_i + \sum_{i=1}^G\zbf_i^H\Hbf_i\Dbf_{\lambdabf_i}\Hbf_i^H\zbf_i \nn\\
&-\sum_{i=1}^G\zbf_i^H\Hbf_i\Dbf_{\lambdabf_i}\Hbf_i^H\Rbf_{i^-}^{-1}(\lambdabf)\Hbf_i\Dbf_{\lambdabf_i}\Hbf_i^H\zbf_i \nn\\
=&\sum_{i=1}^G\zbf_i^H\Hbf_i\Dbf_{\lambdabf_i}\Hbf_i^H(\Ibf-\Rbf_{i^-}^{-1}(\lambdabf)\Hbf_i\Dbf_{\lambdabf_i}\Hbf_i^H)\zbf_i \nn \\[-.5em]
&+\sigma^2\sum_{i=1}^G\lambdabf_i^T\gammabf_i. \label{SCA:g}
\end{align}
From  the optimal  $\wbf_i^o$ expression in \eqref{w_o}, based on the definition of $\alpha_{ik}^o$, we have $\alphabf_i^o = \Dbf_{\lambdabf_i^o}\Hbf_i^H\wbf_i^o$. Substituting this expression into \eqref{w_o}, we have
\begin{align*} 
\wbf_i^o &= \Rbf_{i^-}^{-1}(\lambdabf^o)\Hbf_i\Dbf_{\lambdabf_i^o}\Hbf_i^H \wbf_i^o, \ i\in \Gc,
\end{align*}
or equivalently,
\begin{align}\label{lambda_opt2}
\left(\Ibf -\Rbf_{i^-}^{-1}(\lambdabf^o)\Hbf_i\Dbf_{\lambdabf_i^o}\Hbf_i^H\right)\wbf_i^o = {\bf 0}.
\end{align}

As explained in the proof of Theorem~\ref{thm1}, by the iterative procedure of the SCA method, if $\zbf_i$ converges to the optimum $\zbf_i \to \wbf_i^o$ in \eqref{SCA:g}, the optimal $\lambdabf^\star(\zbf)$ for the dual problem $\Dc_\SCA(\zbf)$ also converges  $\lambdabf^\star(\zbf) \to \lambdabf^o$. By  \eqref{lambda_opt2}, it follows that
\begin{align*}
\max_{\lambdabf} g(\wbf^o,\lambdabf)&=\sigma^2\sum_{i=1}^G{\lambdabf_i^o}^T\gammabf_i=\sigma^2{\lambdabf^o}^T\gammabf.
\end{align*}
As $\zbf\to \wbf^o$,  $\Dc_\SCA(\zbf)\to \Dc_\SCA(\wbf^o)$, $\Pc_\SCA(\zbf) \to \Pc_o$, and we obtain the minimum objective value of $\Pc_o$  as the above. 
\endIEEEproof

\section{Proof of Proposition~\ref{prop2}}\label{app:prop2}
\IEEEproof
We first present the following lemma regarding the average of random variables.
\begin{lemma}\label{lemma0}
Suppose $\{x_n\}$ is a  sequence of i.i.d. random variables with $E(|x_n|)< \infty$, $E(x_n)=0$, and $\{c_n\}$ is a bounded sequence, where $c_n$ is real, for all $n$.   Then, $\frac{1}{N}\sum_{n=1}^Nc_nx_n \to 0$ almost surely (a.s.) as $N \to \infty$.
\end{lemma}
\IEEEproof
The result deals with an independent but not identically distributed sequence of random variables. It can be viewed as a variation of Kolmogorov's Strong Law of Large Numbers (SLLN).  The proof of this lemma follows the similar steps in the proof of SLLN, and thus is omitted here. The proof of SLLN can be found in \cite[Theorem 7.5.1]{Resnick:book}.
\endIEEEproof
Using Lemma~\ref{lemma0}, we have the following result.
\begin{lemma}\label{lemma1}
 Consider  any two independent   channel vectors $\hbf_{ik}$ and $\hbf_{il}$, $l\neq k$, each containing i.i.d. zero-mean  elements. Then
\begin{align}\label{prop2_eqn}
\lim_{N\to \infty}\frac{1}{N}\hbf_{ik}^H\Rbf^{-1}(\lambdabf)\hbf_{il}=0 \ \ \text{a.s.}
\end{align}
\end{lemma}

\IEEEproof
From the expression of $\Rbf(\lambdabf)$ in \eqref{R}, define
\begin{align}
\Rbf(\lambdabf;{ik}^-)&\triangleq \Rbf(\lambdabf)-\lambda_{ik}\gamma_{ik}\hbf_{ik}\hbf_{ik}^H.
\end{align}
To simplify the notation, let $\rho_{ik}\triangleq \lambda_{ik}\gamma_{ik}$. Using the formula $(\Abf + \ubf\ubf^H)^{-1} = \Abf^{-1} - \frac{\Abf^{-1}\ubf\ubf^H\Abf^{-1}}{1+\ubf^H\Abf^{-1}\ubf}$, where $\Abf$ is an  $n\times n$ matrix and  $\ubf$ is an $n\times 1$ vector, we have
\begin{align}
&\hspace*{-1em}\hbf_{ik}^H\Rbf^{-1}(\lambdabf)\hbf_{il} = \hbf_{ik}^H\left(\Rbf(\lambdabf;{ik}^-)+\rho_{ik}\hbf_{ik}\hbf_{ik}^H\right)^{-1}\hbf_{il} \nn\\
&=\hbf_{ik}^H\Rbf^{-1}(\lambdabf;{ik}^-)\hbf_{il}\Big(\!1\!-\!\frac{\rho_{ik}\hbf_{ik}^H\Rbf^{-1}(\lambdabf;{ik}^-)\hbf_{ik}}{1+\rho_{ik}\hbf_{ik}^H\Rbf^{-1}(\lambdabf;{ik}^-)\hbf_{ik}}\!\Big) \nn\\
&= \frac{\hbf_{ik}^H\Rbf^{-1}(\lambdabf;{ik}^-)\hbf_{il}}{1+\rho_{ik}\hbf_{ik}^H\Rbf^{-1}(\lambdabf;{ik}^-)\hbf_{ik}}, \quad \forall k,l,i. \label{prop2_eqn2}
\end{align}

For $l\neq k$, define $\Rbf(\lambdabf;(ik,il)^-)\triangleq \Rbf(\lambdabf;{ik}^-)-\rho_{il}\hbf_{il}\hbf_{il}^H$. Using $\Rbf(\lambdabf;(ik,il)^-)$, we apply the  same procedure above to $\hbf_{ik}^H\Rbf^{-1}(\lambdabf;{ik}^-)\hbf_{il}$ in \eqref{prop2_eqn2} again and obtain \eqref{prop2_eqn3}.
Note that, by our construction, $\Rbf(\lambdabf;(ik,il)^-)$ at the right-hand side (RHS) of \eqref{prop2_eqn3} is no longer a function of $\hbf_{ik}$ or $\hbf_{il}$.
\begin{figure*}[!t]
\begin{align}\label{prop2_eqn3}
\hbf_{ik}^H\Rbf^{-1}(\lambdabf)\hbf_{il}
=& \frac{\hbf_{ik}^H\Rbf^{-1}(\lambdabf;(ik,il)^-)\hbf_{il}}{(1+\rho_{ik}\hbf_{ik}^H\Rbf^{-1}(\lambdabf;{ik}^-)\hbf_{ik})(1+\rho_{il}\hbf_{il}^H\Rbf^{-1}(\lambdabf;(ik,il)^-)\hbf_{il})}, \quad l\neq k.
\end{align}
\hrulefill \vspace*{-1.5em}
\end{figure*}

 Let $\Rbf^{-1}(\lambdabf;(ik,il)^-)=\Ubf\Deltabf\Ubf^H$, where $\Deltabf$ is a diagonal matrix containing the eigenvalues $\{\delta_n\}$ of  $\Rbf^{-1}(\lambdabf;(ik,il)^-)$. We have\\[-2em]
\begin{align}\label{prop2_eqn5}
\hspace*{-.6em}\hbf_{ik}^H\Rbf^{-1}(\lambdabf;(ik,il)^-)\hbf_{il}
& =\tilde{\hbf}_{ik}^H\Deltabf\tilde{\hbf}_{il} = \! \sum_{n=1}^N\delta_n\tilde{h}_{ik,n}^*\tilde{h}_{il,n}
\end{align}
where $\tilde{\hbf}_{ik}=[\tilde{h}_{ik,1},\ldots,\tilde{h}_{ik,N}]^T\triangleq\Ubf^H\hbf_{ik}$, and $\tilde{\hbf}_{il}$ is similarly defined. For $\hbf_{ik}$ and $\hbf_{il}$ being independent and zero-mean, $\tilde{\hbf}_{ik}$ and $\tilde{\hbf}_{il}$ are also independent and zero-mean. Let $x_n =\tilde{h}_{ik,n}^*\tilde{h}_{il,n}$. We have $E(x_n)=E(\tilde{h}_{ik,n}^*)E(\tilde{h}_{il,n})=0$. Since $\lambda_{i,k}\ge 0$, from the structure of $\Rbf(\lambdabf;(ik,il)^-)$, it is easy to see that    $0<\delta_n\le 1$,  $\forall n$. Thus, the sequence $\{\delta_n,1\le n \le N\}$ is bounded (for any given $N$). By Lemma~\ref{lemma0}, we have $\frac{1}{N}\sum_{n=1}^N\delta_n\tilde{h}_{ik,n}^*\tilde{h}_{il,n} \to 0$ a.s., or equivalently,
\begin{align}
\lim_{N\to \infty}\frac{1}{N}\hbf_{ik}^H\Rbf^{-1}(\lambdabf;(ik,il)^-)\hbf_{il}
& =0 \ \ \text{a.s.}
\end{align}
 Applying this to \eqref{prop2_eqn3}, we have \eqref{prop2_eqn}, for $l\neq k$.
\endIEEEproof

By  \eqref{prop2_eqn2}, we have
\begin{align}\label{prop2_eqn4}
\hbf_{ik}^H\Rbf^{-1}(\lambdabf)\hbf_{ik}&= \frac{\hbf_{ik}^H\Rbf^{-1}(\lambdabf;{ik}^-)\hbf_{ik}}{1+\rho_{ik}\hbf_{ik}^H\Rbf^{-1}(\lambdabf;{ik}^-)\hbf_{ik}}.
\end{align}
Similar to \eqref{prop2_eqn5}, by decomposing $\Rbf^{-1}(\lambdabf;ik^-)$, we have
\begin{align}\label{prop2_eqn6}
\frac{1}{N}\hbf_{ik}^H\Rbf^{-1}(\lambdabf;ik^-)\hbf_{ik}
& =\frac{1}{N}\sum_{n=1}^N\delta_n'|\tilde{h}_{ik,n}'|^2
\end{align}
where $\Rbf^{-1}(\lambdabf;ik^-)=\Ubf'\Deltabf'\Ubf'^H$ with $\Deltabf'\triangleq\diag([\delta'_1,\ldots,\delta'_N])$,  and $\tilde{\hbf}_{ik}'\triangleq \Ubf'^H\hbf_{ik}$ with $\tilde{h}'_{ik,n}$ being the $n$th element in $\tilde{\hbf}_{ik}'$. Since elements in $\hbf_{ik}$ are i.i.d.,   $\tilde{h}'_{ik,n}$'s are  i.i.d. Also similarly, we have $0<\delta'_n\le 1$,  $\forall n$, for   $\Rbf^{-1}(\lambdabf;ik^-)$. Denote $E(|\tilde{h}_{ik,n}'|^2)\triangleq\sigma_{h,ik}^2$ and define $x_n \triangleq |\tilde{h}_{ik,n}'|^2-\sigma_{h,ik}^2$.  It is easy to verify that $\{x_n\}$ and $\{\delta'_n\}$ satisfy the conditions in Lemma~\ref{lemma0}, and thus we have, as $N\to \infty$,  $\frac{1}{N}\sum_{n=1}^N\delta_n'(|\tilde{h}_{ik,n}'|^2-\sigma_{h,ik}^2)\to 0~ \text{a.s.}$, or equivalently,
\begin{align}\label{prop2_eqn7}
\frac{1}{N}\hbf_{ik}^H\Rbf^{-1}(\lambdabf;ik^-)\hbf_{ik} -\frac{\sigma_{h,ik}^2}{N}\sum_{n=1}^N\delta_n' \to 0 \ \ \text{a.s.}
\end{align}
Applying this to  \eqref{prop2_eqn4}, we have $\lim_{N\to \infty}\frac{1}{N} \hbf_{ik}^H\Rbf^{-1}(\lambdabf)\hbf_{ik} =c ~~\text{a.s.}$, for some $c>0$.\footnote{It can be shown by contradiction; Otherwise,  \eqref{lambda_eqn1} would not hold.}

For $\lambda_{ik}$ being the solution of \eqref{lambda_eqn1},  by Lemma~\ref{lemma1} and the above result, we have
\begin{align*}
&\lambda_{ik} (1+\gamma_{ik})\hbf_{ik}^H\Rbf^{-1}(\lambdabf)\hbf_{il} \nn\\ &=
\frac{\hbf_{ik}^H\Rbf^{-1}(\lambdabf)\hbf_{il}}{\hbf_{ik}^H\Rbf^{-1}(\lambdabf)\hbf_{ik}} 
=\frac{\hbf_{ik}^H\Rbf^{-1}(\lambdabf)\hbf_{il}/N}{\hbf_{ik}^H\Rbf^{-1}(\lambdabf)\hbf_{ik}/N} \to 0 \ \ \text{a.s.}
\end{align*}
as $N\to \infty$, for $\forall k,l\in \Kc_i$ and $l\neq k$.
Thus, the second equation in \eqref{lambda_eqn}  asymptotically holds, and  \eqref{lambda_mtx_eqn} asymptotically holds. Since \eqref{lambda_mtx_eqn} is a sufficient condition for \eqref{lambda_opt1}, it follows that \eqref{lambda_opt1} also asymptotically holds.  In other words, the solution $\lambdabf$ of \eqref{lambda_eqn1} converges to the solution of \eqref{lambda_opt1}  almost surely.
\endIEEEproof

\section{Proof of Proposition~\ref{prop3}}\label{app:prop3}

\IEEEproof
From    \eqref{lambda_eqn1} and  \eqref{prop2_eqn4},  we have, for $\forall \ k\in\Kc_i, i\in \Gc$,
\begin{align}
\frac{\lambda_{ik} (1+\gamma_{ik})\hbf_{ik}^H\Rbf^{-1}(\lambdabf;{ik}^-)^{-1}\hbf_{ik}}{1+\rho_{ik}\hbf_{ik}^H\Rbf^{-1}(\lambdabf;{ik}^-)^{-1}\hbf_{ik}}=1.
 \label{prop3_eqn1} 
\end{align}
Substituting $\rho_{ik}=\lambda_{ik}\gamma_{ik}$ back into \eqref{prop3_eqn1}, bringing the denominator to the RHS, and removing the common terms at both sides, we have
\begin{align}\label{prop3_eqn2}
\lambda_{ik}=\frac{1}{\hbf_{ik}^H\Rbf^{-1}(\lambdabf;{ik}^-)^{-1}\hbf_{ik}}.
\end{align}
Following \eqref{prop2_eqn6},   we note that $\tilde{\hbf}'_{ik}$ and $\hbf_{ik}$ have the same distribution. Thus, for $\hbf_{ik}=\sqrt{\beta_{ik}}\gbf_{ik}$ with $\gbf_{ik}\sim \Cc\Nc({\bf 0},\Ibf)$,  we have $E(|\tilde{h}_{ik,n}'|^2)=\beta_{ik}$.
By  \eqref{prop2_eqn7} and the fact that  $\sum_{n=1}^N\delta'_n=\tr(\Deltabf')=\tr(\Rbf^{-1}(\lambdabf;ik^-))$, we have, as $N\to \infty$,
\begin{align}\label{prop3_eqn3}
\hspace*{-.5em}\frac{1}{N}\hbf_{ik}^H\Rbf^{-1}(\lambdabf;ik^-)\hbf_{ik} - \frac{\beta_{ik}}{N}\tr(\Rbf^{-1}(\lambdabf;ik^-)) \to 0 \  \text{a.s.}
\end{align}
From \eqref{prop3_eqn2} and \eqref{prop3_eqn3}, it follows that, as $N\to \infty$,
\begin{align*}
\frac{\frac{1}{N}}{\frac{1}{N}\hbf_{ik}^H\Rbf^{-1}(\lambdabf;{ik}^-)^{-1}\hbf_{ik}} -\frac{\frac{1}{N}}{ \frac{\beta_{ik}}{N}\tr(\Rbf^{-1}(\lambdabf;ik^-))} \to 0 \ \ \text{a.s.}
\end{align*}
Thus, we have,   as $N\to \infty$,
\begin{align}\label{prop3_eqn4}
\lambda_{ik}\beta_{ik} &- \frac{1}{\tr(\Rbf^{-1}(\lambdabf;ik^-))} \to 0 \ \ \text{a.s.}
\end{align}
By Jensen's inequality,  $\frac{1}{\frac{1}{N}\tr(\Rbf^{-1}(\lambdabf;ik^-))}\le \frac{1}{N}\tr(\Rbf(\lambdabf;ik^-))$. For $\hbf_{ik}=\sqrt{\beta_{ik}}\gbf_{ik}$, we have
\begin{align}
 \frac{\frac{1}{N}}{\frac{1}{N}\tr(\Rbf^{-1}(\lambdabf;ik^-))}&\le\frac{1}{N^2}\tr(\Rbf(\lambdabf; ik^-)) \nn\\
=\frac{1}{N}&\Big[1+\mathop{\sum\sum}_{jl\neq ik}\lambda_{jl}\beta_{jl}\gamma_{jl}\Big(\frac{1}{N}\sum_{n=1}^N|g_{jl,n}|^2\Big)\Big] \nn
\end{align}
Note by Jensen's inequality that, the gap between two sides of the above inequality reduces when the difference between the diagonal elements of $\frac{1}{N}\Rbf(\lambdabf; ik^-)$ reduces\footnote{For convex function $f(x)$, as the range of $x$ becomes smaller, $f(x)$ is  closer to a linear function.}. In this case, the above bound becomes tight, and  inequality becomes equality. Examining the diagonal elements, $[\frac{1}{N}\Rbf(\lambdabf; ik^-)]_{nn}=\frac{1}{N}(1+\mathop{\sum\sum}_{jl\neq ik}\lambda_{jl}\beta_{jl}\gamma_{jl}|g_{jl,n}|^2)$, for all $n$, we verify that they have diminishing variance among them as $N\to \infty$. It follows that, for \eqref{prop3_eqn4},  as $N\to \infty$,
\begin{align*}
\lambda_{ik}\beta_{ik}- \frac{1}{N}(1+\mathop{\sum\sum}_{jl\neq ik }\lambda_{jl}\beta_{jl}\gamma_{jl}) \to 0 \ \ \text{a.s.}
\end{align*}
Let $\bar{\lambda}_{ik} \triangleq \lambda_{ik}\beta_{ik}$. Rewrite the above,  we have, as $N\to \infty$,
\begin{align}\label{prop3_eqn5}
\Big(1+\frac{\gamma_{ik}}{N}\Big)\bar{\lambda}_{ik}
\to \frac{ 1+\sum_{j=1}^G\sum_{l=1}^{\Kc_j}\bar{\lambda}_{jl}\gamma_{jl}}{N}  \ \ \text{a.s.} 
\end{align}
for  $\forall k\in \Kc_i$, $i\in \Gc$, where the RHS limit is the same for all $ k\in \Kc_i$, $i\in \Gc$. It follows that for $k\in \Kc_i, l\in \Kc_j,i,j\in \Gc$, $ik\neq jl$, as $N\to \infty$,
\begin{align*}
\frac{\bar{\lambda}_{jl}}{\bar{\lambda}_{ik}}=\frac{1+\gamma_{ik}/N}{1+\gamma_{jl}/N}=1+\Oc\Big(\frac{1}{N}\Big).
\end{align*}
Substituting $\bar{\lambda}_{jl}=\bar{\lambda}_{ik}\big(1+\Oc\big(\frac{1}{N}\big)\big)$, for $jl\neq ik$, into the RHS of \eqref{prop3_eqn5}, we have, as $N\to \infty$,
\begin{align*}
\bar{\lambda}_{ik}\! &= \frac{1}{\displaystyle N\!-\!\mathop{\sum\sum}_{jl\neq ik}\gamma_{jl}-\Oc\Big(\frac{1}{N}\Big)} \!= \frac{1}{\displaystyle N\!-\!\mathop{\sum\sum}_{jl\neq ik}\gamma_{jl}}+o\Big(\!\frac{1}{N^2}\!\Big) \end{align*}
which is \eqref{prop3:eqn0}.
\endIEEEproof

\vspace*{-.8em}

\bibliographystyle{IEEEtran} 
\bibliography{IEEEabrv,\bibhome/master,\bibhome/Book,mybib}

\end{document}